\documentclass[fleqn,usenatbib]{mnras}

\usepackage{newtxtext,newtxmath}

\usepackage{caption}
\usepackage{subcaption}

\usepackage[T1]{fontenc}

\DeclareRobustCommand{\VAN}[3]{#2}
\let\VANthebibliography\thebibliography
\def\thebibliography{\DeclareRobustCommand{\VAN}[3]{##3}\VANthebibliography}

\usepackage{graphicx}	
\usepackage{amsmath}	

\usepackage{placeins}
\usepackage{orcidlink}

\title[ArkenstoneBH]{ArkenstoneBH. A model for high-specific energy black hole feedback in cosmological simulations}

\author[J. Sullivan et al.]{
\parbox[t]{\textwidth}{
James M. Sullivan$^{1}$\thanks{E-mail: jms2561@columbia.edu}\orcidlink{0009-0008-0904-5535},
Greg L. Bryan$^{1}$\orcidlink{0000-0003-2630-9228},
Matthew C. Smith$^{2}$\orcidlink{0000-0002-9849-877X},
Jake S. Bennett$^{3}$\orcidlink{0000-0002-8573-2993},
Drummond B. Fielding$^{4}$\orcidlink{0000-0003-3806-8548},
Bryan A. Terrazas $^{5}$\orcidlink{0000-0001-5529-7305},
Sophie Koudmani $^{6,7}$\orcidlink{0000-0002-1528-5091},
Rachel S. Somerville $^{8}$\orcidlink{0000-0002-6748-6821},
Michaela Hirschmann $^{9}$\orcidlink{0000-0002-3301-3321}
}
\\
$^{1}$Department of Astronomy, Columbia University, 550 West 120th Street, New York, NY, 10027, USA\\
$^{2}$Max-Planck-Institut für Astrophysik, Karl-Schwarzschild-Str. 1, D-85748, Garching, Germany\\
$^{3}$Center for Astrophysics | Harvard $\&$ Smithsonian, 60 Garden Street, Cambridge, MA 02138, USA\\
$^{4}$Department of Physics, New York University, New York, NY 10003, USA\\
$^{5}$Department of Physics $\&$ Astronomy, Oberlin College, Oberlin, OH 44074, USA \\
$^{6}$Centre for Astrophysics Research, Department of Physics, Astronomy and Mathematics, University of Hertfordshire, College Lane, Hatfield, AL10 9AB, UK\\
$^{7}$Kavli Institute for Cosmology, University of Cambridge, Madingley Road, Cambridge, CB3 0HA, UK\\
$^{8}$Center for Computational Astrophysics, Flatiron Institute, 162 Fifth Avenue, New York, NY 10010, USA\\
$^{9}$Institute of Physics, Laboratory for Galaxy Evolution, EPFL, Observatory of Sauverny, Chemin Pegasi 51, CH-1290 Versoix,
Switzerland
}

\date{Accepted XXX. Received YYY; in original form ZZZ}

\pubyear{2025}

\begin{document}
\label{firstpage}
\pagerange{\pageref{firstpage}--\pageref{lastpage}}
\maketitle

\begin{abstract}
AGN feedback is a key piece of galaxy evolution but is difficult to model due to its high specific energies, multiphase nature, and limited simulation resolutions. \textsc{Arkenstone} is a subgrid framework for representing multiphase flows in coarse resolution simulations that has been used to model stellar feedback driven galactic winds. It ensures the correct treatment of high specific energy feedback that would otherwise be challenging to model accurately in Lagrangian simulations. We introduce the new \textsc{Arkenstone BH} model, which extends the \textsc{Arkenstone} framework to model black hole feedback. We focus on describing the first piece of this framework, which follows the hot, high specific energy phase of these outflows. The second piece, which treats their multiphase structure with a scheme for modeling unresolved cold clouds,  will be implemented and described in a later paper. We present \textsc{Arkenstone BH} in simulations of an isolated galaxy to demonstrate the framework and its ability to capture high specific energy feedback that interacts only weakly with cold, dense gas. We show how these energetic outflows suppress star formation in our isolated galaxy by counteracting the inflow of gas from the circumgalactic medium into the interstellar medium. This work is part of the `Learning the Universe' collaboration, which aims to understand the Universe's underlying physics and initial conditions. 
\end{abstract}





\section{Introduction} \label{sec:intro}

Supermassive black holes are found at the center of nearly all galaxies. They are believed to coevolve with their host galaxies, growing alongside the galaxy's stellar bulge. As they accrete gas, they power active galactic nuclei (AGN), releasing energy in the form of radiation and fast moving particles. This feedback is widely invoked to regulate galaxy growth, quench massive galaxies ($M_\text{halo} \gtrsim 10^{12} \, \rm M_\odot$), reproduce the observed galaxy stellar mass function, and suppress the gas accretion that fuels star formation \citep[see reviews by][]{Kormendy13, Eckert21}.

To quench galaxies, this feedback must offset radiative cooling and limit the inflow of cool gas from the intergalactic medium (IGM) through the circumgalactic medium (CGM) and into the interstellar medium (ISM). Simulations that include only radiative cooling, star formation, and gravity significantly overpredict stellar masses and baryon fractions \citep{White91,Keres09, Choi15, Li15, AnglesAlcazar17, Su19a}. In less massive halos ($M_\text{halo} \lesssim 10^{12} \, \rm M_\odot$), stellar feedback can regulate star formation, but it is not powerful enough on its own to quench more massive halos \citep[e.g.,][]{Somerville15}.

In higher mass halos, however, AGN feedback is the leading candidate for suppressing cooling flows and quenching star formation. Cosmological simulations that include AGN produce more realistic baryon fractions and reproduce key black hole (BH)-galaxy scaling relations \citep{Sijacki07, Booth13, Teyssier11, Martizzi12, Vogelsberger13, Barai14}, in agreement with semi-analytical models \citep{Bower06, Croton06, DeLucia07, Somerville08, Guo11}. 

AGN feedback is commonly classified into two modes: a radiatively efficient and a radiatively inefficient mode. The former, also called quasar mode feedback, typically occurs in BHs accreting near the Eddington rate, the maximum steady accretion rate at which radiation pressure balances gravitational infall. This is most common when AGN are surrounded by large reservoirs of cool, molecular gas \citep{Fabian12}. In this mode, winds emanating from the accretion disk may efficiently couple with the surrounding ISM to decrease the available cool gas mass. In the case that they do not couple efficiently with the ISM, they may still deposit significant amounts of energy into the CGM and IGM \citep{Heckman14}. However, the extent to which this occurs and its impact on galaxy gas content remains uncertain.

The second mode, radiatively inefficient feedback, typically occurs in BHs accreting far below the Eddington limit. It is characterized by collimated outflows or jets and is often (but not always) associated with massive galaxies that host hot gaseous halos or reside at the centers of groups and clusters \citep{Fabian12, Heckman14}. Extended cavities and bubbles in the hot gas of both clusters and individual galaxies revealed by X-ray observations \citep{Blanton2001, Fabian2003, HlavacekLarrondo15} spatially coincide with AGN-driven radio emission, providing strong evidence that they are inflated by jets \citep[see review by][]{Eckert21}. 

Across (and within) each of the two modes discussed above, AGN feedback produces multiphase outflows, containing ionized, neutral, and molecular gas. These different phases have been traced by X-ray, UV, optical and infrared absorption lines, helping to probe outflows from the galaxy center at sub-pc scales out to the larger scales within the CGM. These observations have also revealed that these outflows span a wide range of temperatures and velocities. Quasar mode outflows can range in temperature from $10^{1-2}\, \rm K$ for the molecular gas to $10^{6-7}\, \rm K$ for the highly ionized gas \citep{Cicone18}. Similarly, these outflows have a wide range of measured velocities ($10^{2-4}  \rm km \, s^{-1}$) \citep{Chartas07, Cattaneo09, Seebeck24, Xu25}. Radiatively inefficient AGN jets can similarly drive outflows at the upper end of this velocity range, with observed jet propagation speeds reaching at $v \gtrsim 0.1c$ \citep{Odea09}. They additionally inflate cavities filled with relativistic plasma while simultaneously driving shocks into the surrounding gas, heating it to temperatures of $T \gtrsim 5 \, \rm keV$ \citep{Snios18, Gilli19}.

The different phases of feedback produce distinct effects on the host galaxy, depending on the balance of mass and energy. To capture these multiphase outflows in simulations, it is useful to categorize feedback into two types: ejective (low specific energy) and preventative (high specific energy). In the former, low specific energy winds carry large amounts of material from the ISM out into the CGM and IGM. This consequently reduces the availability of gas to form stars and fuel subsequent BH accretion. However, this ejected mass can fall back into the ISM in some instances, helping fuel future star formation. This depends both on whether or not ejected gas can escape the galaxy's gravitational potential and also on how effectively the outflow transfers kinetic and thermal energy to the surrounding gas \citep{Terrazas20}.  

In contrast, high specific energy outflows (in either wind or jet form) act primarily as preventative feedback, carrying significant energy but only small amounts of mass into the CGM and beyond. By heating the surrounding gas, they can counter radiative cooling and suppress the inflow of cool gas that fuels star formation \citep{Begelman04, Springel05, Hopkins06a, Hopkins06b, Hopkins07, Hopkins08, Johansson09, Hopkins10, Ostriker10, Faucher12, Dubois13, Barai14, Weinberger17a, Pillepich18, Richings18a, Richings18b, Torrey20, Farcy25}. The correct balance of ejective versus preventative feedback, however, remains uncertain.

Correctly modeling both of these feedback types is thus a crucial area of computational astrophysics. However, modeling AGN feedback covers a wide range of scales, from the innermost stable orbit at roughly twice the Schwarzschild radius ($r_\text{g} \sim 10^{-4}\,\rm pc$ for a $10^9 \rm \, M_\odot$ BH), to the Bondi radius ($\sim 10-100 \, \rm pc$), and well into the CGM and beyond, where radiative cooling must be offset (several 100 kpc or more). Resolving all of these scales together is therefore extremely difficult. Simulations including general-relativistic magnetohydrodynamics (GRMHD) have modeled the creation and evolution of these outflows in simulations resolving the black hole event horizon and accretion disk \citep{Hawley04, Tchekhovskoy11, McKinney12, White19}. However, most cannot simulate beyond about $10^{4}$ times the gravitational radius, remaining sub-pc in scale \cite[although see very recent work by][]{Cho23, Cho24, Hopkins24a, Hopkins24b, Hopkins24c, Guo2025}. Larger scale simulations, in which galaxy and AGN populations can be studied, are unable to reach the sub-parsec resolutions required by these models. This creates a need for subgrid AGN feedback models.

Subgrid models for AGN feedback are included in most modern cosmological simulations \citep{Schaye15, Weinberger18, Dave19, Costa20, Schaye23, Farcy25}\footnote{EAGLE, IllustrisTNG, SIMBA, BOLA, FLAMINGO, and MISTRAL}, though there are substantial differences across these simulations in how the models work (see our discussion in Sec. \ref{subsec:ModelComparisons}). However, we identify the high specific energy, preventative mode of feedback and the ability to produce and evolve multiphase outflows as a key focus for advancement of these models.

As previously described, jets and other high specific energy AGN outflows can inject enormous amounts of energy into their surroundings, often exceeding the typical energy output of radiative winds \citep{Merloni2007}. These energetic flows are likely to play a key role in regulating the thermodynamic state of gas in massive halos, offsetting radiative cooling, shaping X-ray observables, and suppressing star formation. Accurately capturing the energetics and impact of these high specific energy outflows is therefore essential for cosmological simulations aiming to model the evolution of massive galaxies.

Simulating these high specific energy outflows presents a significant challenge in cosmological simulations due to their refinement strategies and limited resolution. Most use Lagrangian or quasi-Lagrangian refinement schemes in which gas cells maintain approximately constant mass, meaning the resolution is concentrated in dense material. Jets, however, deposit large amounts of energy into low density, low resolution regions and do not significantly increase the resolution due to their low mass loading. They thus tend to be poorly resolved by such methods, and therefore their interactions with the ambient gas may be inaccurate \citep{Bourne17, Weinberger17b, Weinberger23}. The mass and energy input could then become temporally and spatially under-resolved (see discussion in Sec.~\ref{subsec:ArkRecoupling}~$\&$~\ref{subsec:ArkRefinement}).

The temporal resolution problem arises from injecting feedback energy in discrete events. For a fixed energy input rate, matching the high specific energies of jets requires energy to accumulate until it can raise a fixed mass Lagrangian cell to the target specific energy. This introduces artificial burstiness, as energy is deposited intermittently rather than continuously. Lower resolution exacerbates this effect by requiring larger, less frequent events. The problem is further compounded when sampling a distribution of temperatures and velocities to capture the multiphase nature of AGN feedback, which demands substantially higher mass resolution \citep[see discussion in][]{Smith24a}.

The \textsc{Arkenstone} model was developed and implemented in the AREPO code to address this issue for stellar feedback \citep{Smith24a, Smith24b}. In it, winds can be split into several components. Each component is given separate mass and energy loading. These loadings then help to set the outflow properties (Section \ref{subsubsec:ArkBH}). The components themselves are split up into particles which share a launch velocity and temperature. These can in principle be calibrated to results from small scale simulations \cite[e.g.][]{Kim20}. We refer to these particles simply as `wind particles' for consistency, even though this work focuses primarily on jet mode AGN feedback rather than radiative winds from either supernovae or AGN. The \textsc{Arkenstone Hot} scheme specifically targets high specific energy outflows, ensuring their evolution within and impact on the CGM is properly resolved \citep{Smith24a}. The newer \textsc{Arkenstone Cold} scheme models cold clouds embedded in outflows using `cloud particles' \citep{Smith24b}. This model captures the exchange of energy, mass, and momentum between the clouds and surrounding hot medium. Across both schemes, the particles are uncoupled at launch and then injected just outside the ISM. This gives greater control over the wind properties and avoids effects tied to the winds propagating through a poorly resolved ISM. Injecting the winds at the source and trying to resolve their evolution in these lower resolution simulations would produce an incorrect evolution of the energy and mass loadings. In summary, the method allows multiphase stellar outflows and their interactions within the CGM to be resolved in cosmological simulations without having to artificially increase the simulations' overall resolution.

In this paper, we present \textsc{Arkenstone BH} (\textsc{ArkBH}), a new extension of the \textsc{Arkenstone} model designed to resolve high specific energy AGN feedback in cosmological simulations. This work is a part of the `Learning the Universe' collaboration\footnote{https://learning-the-universe.org} and supports one of its key goals: improving models of BH feedback. The primary aim of \textsc{ArkBH} is to model both the evolution of AGN driven outflows and their interaction with the CGM, thereby improving our understanding of how preventative feedback regulates gas accretion and suppresses star formation. \textsc{ArkBH} will be later divided into a `Hot' and `Cold' framework, analogous to the stellar implementation of \textsc{Arkenstone}. This will allow us to more fully capture multiphase outflows by modeling the evolution of cold gas within hot outflows. In this work, however, we focus on the development and behavior of the `Hot' outflow framework only.

Our presentation emphasizes the application of the model to jet mode feedback (although the framework itself is quite general), where the challenges of evolving high specific energy outflows are most pronounced. This informs many of our parameter choices, which we discuss further in Sections~\ref{subsubsec:ArkBH}~and~\ref{subsec:Caveats}. While this study is thus framed in the context of jet mode feedback, the underlying model and numerical methods developed here are also applicable to and can be used to study the other modes of AGN feedback described above. 

To make the scope explicit, the present implementation is a model for hot, preventative AGN feedback that couples primarily outside the dense ISM. It is not intended to represent the full spectrum of AGN feedback, nor to resolve jet/ISM interactions inside the unresolved central disk. Our idealized runs are therefore designed to test whether the method can (i) preserve a hot, low-mass outflow numerically, (ii) couple that outflow to CGM gas, and (iii) alter inflow and star formation at galaxy scales.

Further, the simulations we present are intended to introduce the numerical framework of our model and demonstrate its behavior across a range of the parameter space rather than to establish a definitive AGN feedback prescription. Although our results illustrate the impact of varying model parameters, the idealized nature of our setup precludes drawing strong physical conclusions. A detailed investigation of the model’s influence on galaxy evolution is therefore deferred to future cosmological simulations.

The paper is divided as follows. Section~\ref{sec:SimulationSetup} describes the \textsc{ArkBH} model and the computational setup of our simulations. In Section~\ref{sec:Results}, we present results from implementing \textsc{ArkBH} in an isolated galaxy. Section~\ref{sec:Discussion} compares \textsc{ArkBH} with other subgrid black hole feedback models, details the limitations of our tests, and discusses future applications for \textsc{ArkBH}. Section~\ref{sec:Conclusion} summarizes our results and their implications.

\section{Methodology}
\label{sec:SimulationSetup}

\subsection{\textsc{Arepo} code}
\label{subsec:Arepo Code}
\textsc{Arkenstone} is built on the hydrodynamical code \textsc{Arepo} \citep{Springel10, Pakmor16, Weinberger20}. \textsc{Arepo} uses a finite volume method, solving the equations of hydrodynamics on an unstructured, moving mesh. The mesh is constructed by the Voronoi tessellation of mesh-generating points. These points move with the local fluid velocity, with additional minor corrections to maintain cell regularity. Consequently, this scheme is quasi-Lagrangian, and so can lead to cells losing and gaining mass due to non-zero mass fluxes. However, a (de)refinement scheme is implemented to address the non-zero mass fluxes between cells. Cells are merged or split to constrain them to within a factor of two of the desired mass resolution (the target mass, $m_{\text{target}}$). Additional criteria can be used to further enforce mass or spatial resolution constraints within the simulation. In all runs, we turn on a volume based refinement limit, which restricts neighboring cells to remain within a factor of 8 in volume. Together with the mass based criterion, this defines our default refinement configuration.

\subsection{Physics model}
\label{subsec:Model Physics}

We use the same physical model as the IllustrisTNG simulations \citep{Vogelsberger13}, but exclude magnetic fields as well as both BH and stellar wind feedback (although the \citealt{Springel03} effective equation of state, used throughout our runs, does implicitly include stellar feedback). We do plan to perform magneto-hydrodynamical simulations
using ArkBH in future work. For simplicity however, we restrict the
simulations in this study to hydrodynamics. We include a single run that incorporates the TNG BH feedback (as well as two runs without any BH feedback) in order to provide a comparison against \textsc{ArkBH}. Stellar wind feedback is not included in any of our runs in order to isolate the effects of our BH feedback model. The TNG physics model includes radiative cooling from both primordial species \citep{Cen92,Katz96} and metal lines in the presence of a $z = 0$ UV background \citep{Faucher09}. Radiative cooling below $10^4$ K is not included. Corrections are added for self-shielding in dense gas \citep{Rahmati13}. We use the softer effective equation of state (eEoS) described in \cite{Vogelsberger13} to avoid over pressurizing the ISM. The eEoS models the dense, star forming ISM and captures the effects of unresolved small scale effects (i.e. turbulence, thermal instabilities, etc.) on the ISM. Finally, gravity is implemented using a tree-based algorithm. 

\subsubsection{Black Holes}
\label{subsubsec:BlackHoles}
BHs are represented by collisionless, massive sink particles with accretion and dynamics as implemented in IllustrisTNG \citep{Weinberger17a}. We manually set the BH mass at the start of our runs to a fixed value as described in Section~\ref{subsec:TestGalaxies}. The BH growth is then handled using the prescribed physics in \cite{Vogelsberger13}. This includes the Eddington-restricted, Bondi-Hoyle-Lyttleton accretion rate, which is calculated using the surrounding gas' density and sound speed. In rare cases, the mass actually accreted by the BH in a given timestep can be less than predicted by the Bondi rate if there is not enough gas available to remove in the neighboring cells. The remaining mass is accreted onto the BH over subsequent timesteps. We use the adjusted accretion rate (based on the Bondi rate and adjusted for the actual available mass) in each timestep rather than the Bondi estimate to determine the outflow energy/particle budget (see Sec.~\ref{subsubsec:ArkBH}). This prevents us from ejecting more mass than is accreted across a given timestep. 

We also do not properly resolve the BH dynamics, in particular dynamical friction; instead we use the common fix of repositioning the BH particle to the minimum of the gravitational potential at each timestep. In the \textsc{Arkenstone} runs, we turn off all IllustrisTNG BH feedback (both the thermal and kinetic modes) so that \textsc{ArkBH} produces the only BH feedback. We do not implement a switch between different modes of feedback in our ArkBH routine/runs, including only a single jet mode. We then include a single run with the full Illustris-TNG BH feedback model for comparison.

\subsection{BH Feedback}
\label{subsec:BH Feedback}
We present the two BH feedback models used in this work. The primary focus is \textsc{Arkenstone BH}. We then briefly discuss the TNG BH feedback model, which we include for comparison. 

\subsubsection{\textsc{Arkenstone BH}}
\label{subsubsec:ArkBH}

\begin{table*}
    \centering
    \begin{tabular}{|c|| c c c c c c|}
        \hline
        \multicolumn{7}{|c|}{Arkenstone Run Parameters} \\ 
        \hline\hline
         Run & Galaxy &  $\epsilon_{\text{f}}$ & $v_{\text{w}}$ & $\gamma_{\text{m}}$ & $\text{R}_{\text{rec}}$ & $\rho_{\,\text{rec}}$ \\
        \hline\hline

        Fid1 & 1  & 0.0045 & $3 \times 10^4 \, \rm km/s$ & 0.01 & 10 kpc & n/a\\

        Fid2 & 2 & 0.0045 & $3 \times 10^4 \, \rm km/s$ & 0.01 & 10 kpc & n/a\\

        LowV & 1 & 0.0045 & $\mathbf{1 \times 10^4} \, \rm \textbf{km/s}$ & 0.01 & 10 kpc & n/a\\
        
        HighV & 1 & 0.0045 & $\mathbf{1 \times 10^5} \, \rm \textbf{km/s}$ & 0.01 & 10 kpc & n/a\\
        
        LowEff & 1 & \textbf{0.0009} & $3 \times 10^4 \, \rm km/s$ & 0.01 & 10 kpc & n/a\\
        
        HighEff & 1 & \textbf{0.0225} & $3 \times 10^4 \, \rm km/s$ & 0.01 & 10 kpc & n/a\\
        
        Rec2 & 1 & 0.0045 & $3 \times 10^4 \, \rm km/s$ & 0.01 & \textbf{2 kpc} & n/a\\
        
        Rec5 & 1 & 0.0045 & $3 \times 10^4 \, \rm km/s$ & 0.01 & \textbf{5 kpc} & n/a\\

        Rec20 & 1 & 0.0045 & $3 \times 10^4 \, \rm km/s$ & 0.01 & \textbf{20 kpc} & n/a\\
        
        RecRho & 1 & 0.0045 & $3 \times 10^4 \, \rm km/s$ & 0.01 & n/a & \textbf{0.01} $\rho_\mathrm{crit}$\\

        Low $\gamma_\text{m}$ & 1 & 0.0045 & $3 \times 10^4 \, \rm km/s$ & \textbf{0.001} & 5 kpc & n/a\\

        High $\gamma_\text{m}$ & 1 & 0.0045 & $3 \times 10^4 \, \rm km/s$ & \textbf{0.1} & 5 kpc & n/a\\

        Max $\gamma_\text{m}$ & 1 & 0.0045 & $3 \times 10^4 \, \rm km/s$ & \textbf{1.0} & 5 kpc & n/a\\

    \end{tabular}
    \caption{Parameters used across the thirteen \textsc{ArkBH} runs in this work. This includes the the feedback efficiency ($\epsilon_{\text{f}}$), wind velocity ($v_{\text{w}}$), mass factor ($\gamma_{\text{m}}$), and recoupling radius ($\text{R}_{\text{rec}}$) or density ($\rho_{\text{rec}}$). The kinetic and thermal energy loadings ($\eta_{\text{k}}$ and $\eta_{\text{th}}$), not listed here, are set to 0.357 and 0.643 across all the above runs. These are set to produce an outflow Mach number of 1, as in  \citet{Smith24a}}.
    \label{tab:RunParameters}
\end{table*}

The \textsc{ArkBH} framework adopts the same general principles as \textsc{Arkenstone}'s standard stellar feedback mode but with the feedback tied to the black hole accretion rather than star formation rate (SFR).

We parameterise the \textsc{ArkBH} model using a similar framework to that of \citealt{Choi15} (building on the earlier work of \citealt{Ostriker10} and \citealt{Choi12})  and \cite{Farcy25}. The specific energy of our outflow is fixed and the mass accretion rate, $\dot{M}_\text{BH}$, gives us the energy budget of the outflow. This choice is independent of the outflow mode (`wind' versus `jet'), which is primarily determined by the later choices of the outflow opening angle and particle velocities. The energy budget can then be split up into the kinetic and thermal energy budgets:
\begin{equation}
\dot{E}_{\text{k}} = \eta_\text{k} \epsilon_{\text{f}} \dot{M}_{\text{BH}}c^2
\label{eq:Edotka}
\end{equation}
and 
\begin{equation}
\dot{E}_{\text{th}} = \eta_\text{th} \epsilon_{\text{f}} \dot{M}_{\text{BH}}c^2.
\label{eq:Edotth}
\end{equation}

The feedback efficiency, $\epsilon_\text{f}$, sets the fraction of accretion energy that is injected as feedback. $\eta_\text{k}$ and $\eta_\text{th}$, determine how much of this feedback energy takes the form of kinetic versus thermal energy. We set these values to 0.357 and 0.643 respectively, corresponding to a thermal to kinetic energy ratio of 1.8. This choice of values produces an outflow with Mach number 1. We leave a more detailed exploration of these energy loadings to later work.

We then equate the kinetic energy of the AGN feedback, $\dot{E}_{\text{k}}$, to the kinetic energy of the `wind particles' we use to model it:
\begin{equation}
\dot{E}_{\text{k}} = \eta_{k} \epsilon_{\text{f}} \dot{M}_{\text{BH}}c^2 = \frac{1}{2}\dot{M}_{\text{wind}}v_{\text{w}}^2,
\label{eq:Edotk}
\end{equation}
where $v_{\text{w}}$ is the wind particle velocity and $\dot{M}_{\text{wind}}$ is the feedback's mass outflow rate. Again, we refer to the outflow parameters as `wind particles' to be consistent with the original \textsc{Arkenstone} papers \citep{Smith24a, Smith24b, Bennett24}. The parameter values themselves define whether the outflow represents a `wind' versus a `jet'.

We define the ratio, $\psi$, of our wind's mass flux, $\dot{M}_{\text{wind}}$, to the effective accretion rate of the BH, $\dot{M}_{\text{BH}}$:
\begin{equation}
\psi = 2 \, \eta_k \,\epsilon_{\text{f}}\,c^2/v_w^2 = \dot{M}_{\text{wind}}/\dot{M}_{\text{BH}}.
\label{eq:psi}
\end{equation}
This is a product of the outflow specific energy we set, which is itself determined by our choices of the feedback efficiency, wind velocity, and kinetic energy loading. Increasing $\epsilon_{\text{f}}$ raises $\psi$, leading to a larger fraction of inflowing material being ejected rather than accreted. In contrast, since $\psi \propto v_\text{w}^{-2}$, increasing the wind velocity reduces $\dot{M}_{\text{wind}}$. Thus, higher efficiencies enhance both the mass and energy carried by the outflow, whereas higher velocities concentrate the same energy budget into fewer particles with higher specific energy.

The ratio $\psi$ then allows us parameterize the feedback in terms of the large scale gas inflow/accretion rate (limited by the simulation resolution), $\dot{M}_{\text{BH, inf}}$. We adopt here the Eddington-limited Bondi accretion rate (although we note that ArkBH can work with other accretion prescriptions). The relationship between the outflow mass flux, $\dot{M}_{\text{wind}}$, the black hole accretion rate measured at large scales, $\dot{M}_{\text{BH, inf}}$, and the rate mass accretes onto the black hole, $\dot{M}_{\text{BH}}$, is given as:
\begin{equation}
\dot{M}_{\text{wind}} = \dot{M}_{\text{BH, inf}} - \dot{M}_{\text{BH}} = \dot{M}_{\text{BH, inf}}/(1+\frac{1}{\psi}). 
\label{eq:MassFluxes}
\end{equation}
$\dot{M}_{\text{BH}}$ can then be correspondingly written in terms of the Bondi accretion rate as:
\begin{equation}
    \dot{M}_{\text{BH}} = \dot{M}_{\text{BH, inf}}/(1+\psi).
    \label{eq:EffMdot}
\end{equation}

The energy fluxes from equations~\ref{eq:Edotka} and \ref{eq:Edotth} can also be rewritten in terms of $\dot{M}_{\text{BH, inf}}$:
\begin{equation}
    \dot{E}_{\text{k}} = \eta_\text{k} \epsilon_{\text{f}} \frac{1}{1+2 \, \eta_k \,\epsilon_{\text{f}}\,c^2/v_w^2}\dot{M}_{\text{BH, inf}}c^2
    \label{eq:EdotkBondi}
\end{equation}
and 
\begin{equation}
    \dot{E}_{\text{th}} = \eta_\text{th} \epsilon_{\text{f}} \frac{1}{1+2 \, \eta_k \,\epsilon_{\text{f}}\,c^2/v_w^2}\dot{M}_{\text{BH, inf}}c^2.
    \label{eq:EdotthBondi}
\end{equation}

From these expressions, we see that the energy flux depends on both $\epsilon_{\text{f}}$ and $v_\text{w}$ through their coupled influence on $\psi$. An important, and not immediately obvious, consequence is that the choice of one parameter affects the sensitivity of the energy flux to the other. For fixed $v_\text{w}$, the increase in outflow energy with $\epsilon_{\text{f}}$ becomes less pronounced at high efficiencies, particularly at lower velocities. Conversely, at low efficiencies, increasing $v_\text{w}$ produces only modest changes in the total energy budget. These trends are illustrated in Fig.~1 of \cite{Farcy25}.

The expected number of wind particles launched in each timestep, $n_{\text{w}}$, follows from equations~\ref{eq:MassFluxes} and \ref{eq:EffMdot} and is given by: 

\begin{equation}
    n_{\text{w}} = \dot{M}_{\text{BH, inf}}(\frac{1}{1+\frac{1}{\psi}})\frac{\Delta t}{m_{\text{w}}},
    \label{eq:numparticles}
\end{equation}
where $\Delta t$ is the simulation time-step. The wind particle masses, $m_{\text{w}}$, are set by a mass ratio, $\gamma_{\text{m}} \le 1$, and the target mass of our simulation, $m_{\text{target}}$:
\begin{equation}
m_{\text{w}} = \gamma_{\text{m}}\,m_{\text{target}}.
\end{equation}

We launch a wind particle for every integer value of the expected number. We then choose whether to launch an additional particle, with a probability equal to the float remainder. 
Each wind particle is given a constant input velocity, $v_{\text{w}}$. We use a fiducial value, $ v_{\text{w}}= 3 \times 10^4 \,\rm km\,s^{-1}$, which we vary in Sec.~\ref{subsec:JetParams}. The high velocity ensures a high specific energy across the outflow and is chosen to test the upper range of observed/predicted jet velocities (see Section~\ref{sec:intro}). The specific thermal energies, $u$, are then given by:
\begin{equation}
u = \eta_{\text{th}}\frac{\dot{M}_{\text{BH}}}{\dot{M}_{\text{wind}}}\epsilon_{\text{f}}\,c^2.
\end{equation}

The wind particles are initialized at the black hole location and then launched within a biconical region aligned with the galaxy's z-axis. This is done by assigning them a random polar and azimuthal angle within the cone, which has an opening angle of 0.2 radians, or 11.5 degrees. This opening angle is consistent with values used by other simulations \citep[e.g.][]{Monceau12, Yates23} and within observed values \citep{Pushkarev17}. We will explore the choice of opening angle in future work. The wind particles are then hydro-decoupled at launch and recouple into the CGM, which we discuss in the following section.

The ability to vary the outflow opening angle, velocity, and mass/energy loading makes \textsc{ArkBH} capable of modeling both narrow jets and less collimated `radiative' winds. These parameters can also be set to different values for multiple wind phases. When coupled with \textsc{Arkenstone Cold}, we will be able to use \text{ArkBH} to study multiphase outflows. We leave this to future work.

\subsubsection{Arkenstone Decoupling and Recoupling}
\label{subsec:ArkRecoupling} 

Hydro-decoupling the wind particles at launch is key to \textsc{ArkBH} and sets it apart from many other models. We do this because we are primarily trying to model the impact of jets in the CGM and beyond. The choice to prevent the jet from interacting within the ISM at all is not a statement that jets do not interact with the ISM, although some jets may break out of the galaxy's ISM without coupling significant amounts of energy or mass into it \citep[e.g. ][]{Tanner22}. This is discussed further in Section~\ref{subsec:FutureDirections}. Rather, the poor mass resolution in cosmological simulations makes these interactions impossible to model correctly if we inject our high specific energy wind particles locally around the BH. Without decoupling, energy and momentum are injected into the same set of particles (i.e. the black hole `kernel') which is used to compute accretion properties. This can lead to regulation at the `kernel' level, in which the gas particles near the black hole are kept hot and a low density, while the rest of the halo evolves independently. Indeed, as we will see, the TNG model shows evidence of this kind of self-regulation. 

Further, even if we managed to drive large-scale outflows without decoupling the wind particles, injection into the unresolved ISM would not be able to accurately model the high specific energy outflows (i.e. jets). Depending on the outflow parameters, the ISM could also be significantly disrupted (see Section~\ref{subsubsec: RecouplingDistance}). Studying the impact of jets within the ISM and the generation of subgrid models for these effects is therefore a separate (albeit related) area of study \citep[for an example, see][]{Borodina25}. Wind particles are therefore hydro-decoupled at launch and injected at the ISM/CGM boundary using our recoupling criteria (discussed below). We do not include radiative cooling of the particles while they are decoupled as the long cooling time for our high specific energy flows makes these effects negligible. Finally, we note that \textsc{ArkBH} can be easily combined with complementary subgrid models targeting feedback evolution within the ISM or other feedback modes.

After launch, wind particles travel through the simulation until one of two criteria are met. The first possibility is a density-based criterion, where particles recouple when the background density drops below $\rho_\mathrm{rec}$, which in this paper is 0.01 times the critical density threshold for star formation (i.e. $\rho_\mathrm{crit} = 2.86\times10^{-25} \rm \, g \, cm^{-3}$ as defined in the TNG model). The second is a radial criterion,  which recouples particles after they have traveled a certain ($\text{R}_{\text{rec, fiducial}} = 10 \, \rm kpc$) distance. We use only one of these at a time, and the effect of this choice is investigated in Section~\ref{subsec:NumericalParams}. There is also an additional timer that forces particles to recouple, if they have not already, after 200 Myr, but this is never triggered in this work. We do not test the time based recoupling in isolation, as it is expected to behave in the same way as our radial recoupling criterion. In both cases, the recoupling condition effectively limits the maximum propagation distance of decoupled wind particles.

We then include two recoupling methods, simple and displacement recoupling, which \cite{Smith24a} and \cite{Bennett24} describe in detail. The two methods dictate how wind particles recouple into a host cell, defined as the gas cell containing the wind particle when it is flagged for recoupling. In the simple recoupling method, the wind particle just adds its conserved quantities (i.e. mass, momentum, and energy) to the host cell and is removed from the simulation. However, without additional resolution control, the result is a lower specific energy than desired in the CGM host cell, which may be insufficient to move the gas beyond the peak of the cooling curve near $10^5 \, \rm K$, leading to unphysically high radiative losses in the outflow. To address this issue, we also implement displacement recoupling, a novel method that mitigates the refinement problems described in Section~\ref{sec:intro}. In this method, when a wind particle triggers a recoupling criterion, material from the host cell is distributed to its neighboring cells. The wind particle then adds its contents to the now empty host cell, creating a low mass, high specific energy cell. These are then flagged as “hot wind cells” and set to an enhanced target resolution to prevent de-refinement. This method also retains the original mesh shape which saves computational time. Subsequent wind particles can then recouple into these cells through simple recoupling, which is therefore only applied when the host cell is already a “hot wind cell.”

\subsubsection{Arkenstone Refinement}
\label{subsec:ArkRefinement} 

In order to prevent low mass cells filled with hot wind material from immediately de-refining, we also require a refinement scheme. Motivating this, previous work \citep[e.g.,][]{Smith24a} has shown that resolving high specific energy outflows demands a high spatial resolution, exceeding the standard resolution of the CGM. While the analytical argument was originally developed for stellar feedback (following \citealt{Chevalier85}), the same reasoning applies for BH feedback. BH jets and winds, which can hold higher specific energies than stellar feedback, are therefore expected to be even more poorly resolved at fixed mass resolution in cosmological simulations. To address this, we first track the location of the hot wind material by injecting a passive scalar “dye/tracer” into the host cell when a wind particle recouples. This dye is then advected within the hydrodynamics scheme (exponentially decaying on a timescale $t_0$ = 10 Myr) and sets the cell refinement level.

The initial wind tracer is set as 
\begin{equation}
    f_\text{w, init} = f_\text{w, thresh}e^{\frac{0.2 R_{200}}{v_\text{w}t_0}}.
\end{equation}

It is then also key to relax the refinement as the outflow moves farther out into the CGM. This ensures we do not maintain the high resolution indefinitely. We use a decaying tracer to maintain high resolution in the recoupling region and to gradually relax the resolution to the base value, using the same scheme as \cite{Bennett24}. While the threshold to be considered a hot wind cell is set to $f_\text{w} = 1000$, we allow the tracer to decay further before fully relaxing the refinement back to the target gas mass. We set the wind tracer value at which this occurs, $f_\text{w, ref}$, to 0.1.

The resolution varies with $f_\text{w}$ as follows. Cells with $f_\text{w} >1000$ are capped at $m_{\text{target}}\, \times \,\gamma_\text{m}$. Cells below $f_\text{w, ref}$ are not refined above the simulation target mass. In between, the mass resolution is set to 
\begin{equation}
    m_{\text{target}}\times (1 + \frac{(1-\gamma_\text{m})(f_\text{w}^{-\frac{1}{2}} - f_\text{w, ref}^{-\frac{1}{2}})}{f_\text{w, ref}^{-\frac{1}{2}}-f_\text{w, thresh}^{-\frac{1}{2}}})
\label{eq:MassRes}
\end{equation}
in order to smoothly interpolate between the two limits.

\subsubsection{TNG BH Feedback}
\label{subsubsec:TNG BH Feedback}

We briefly describe the TNG BH feedback model as we will compare it the Arkenstone BH method just described. The TNG model,  which \cite{Weinberger17a} and \cite{Pillepich18} describe in greater detail, includes a thermal feedback mode at high accretion rates and a kinetic mode at lower accretion rates. The accretion threshold separating the two modes is set proportional to the black hole mass and the Eddington accretion rate: $0.002\,(\frac{\text{M}_\text{BH}}{10^8 \text{M}_\odot})^2 \dot{M}_\text{Edd}$. The low accretion mode feedback also shuts off if the density surrounding the black hole drops below 0.01 times the star formation density. The available energy in the high accretion state is parameterized as $\Delta E_\text{high} = \epsilon_{f, \text{high}}\epsilon_r \dot{M}_\text{BH}c^2$ and the available energy in the low accretion state is given by $\Delta E_\text{low} = \epsilon_{f, \text{kin}} \dot{M}_\text{BH}c^2$. We adopt the fiducial values from \cite{Weinberger17a}, setting the radiative efficiency to $\epsilon_r = 0.2$ and the coupling efficiencies to $\epsilon_{f, \text{high}} = 0.1$ and $\epsilon_{f, \text{kin}} = 0.2.$
In the high accretion mode, the energy is input as thermal energy in the area surrounding the BH. In the low accretion mode, it is instead inserted as kinetic energy. In this mode, cells within a set feedback radius are uniformly given a momentum kick in a random direction. In our single TNG run, the BH feedback initially operates in the thermal mode. However, it quickly transitions to the kinetic mode and remains in that state for the rest of the simulation.

The TNG feedback model differs from our \textsc{ArkBH} method, in which we constrain our outflow launch direction. Another key difference is that the Illustris-TNG method does not attempt to modify the resolution or refinement of the simulation to resolve winds. As discussed above, we also include the hydro-decoupling of our wind particles, which TNG includes for stellar feedback but not AGN.

\subsection{Test Galaxy}
\label{subsec:TestGalaxies}

We test the new \textsc{ArkBH} framework in an idealised, isolated galaxy, using a slightly modified version of the m12 galaxy from \cite{Su24}. Each run has a gas mass resolution of $8\times 10^4 \, \rm M_\odot$ and a DM mass resolution of $5\times10^5 \, \rm M_\odot$. The initial m12 galaxy has a total halo mass of $1.5 \times 10^{12} \, \rm M_{\odot}$. It assumes a spherically symmetric and isotropic NFW dark matter profile \citep{Springel99} with scale length 20.4 kpc and a Hernquist stellar bulge profile \citep{Hernquist90} with scale length 1 kpc. The bulge mass is $1.5\times 10^{10} \, \rm M_\odot$. The gas and stellar disks are exponential and rotationally supported and respectively have masses $5\times 10^{10}$ and $5\times10^9\rm\, M_\odot$. Their respective scale lengths are 6 and 3 kpc and both have a scale height of 0.3 kpc. The galaxy's gas temperature is set to pressure equilibrium \citep{Springel00}. The galaxy is also given a beta profile, spherical gas halo in hydrostatic equilibrium (making up our `CGM'). The gas halo mass is $1.5\times 10^{11} \, \rm M_\odot$. It is initially run for at least 50 Myr to stabilize the galaxy. The full initial conditions are described in detail in \cite{Su19a} and \cite{Su20}. For our simulations, a background grid is then added to this set of ICs to make it compatible with Arepo. The simulation box has a side length of 3000 kpc. Finally, the resulting galaxy is run with the full TNG model (excluding BHs and their associated feedback) for $\sim 1.2 \, \rm Gyr$. This does include stellar wind feedback, which is then turned off in our runs. All the runs included in this work begin after this point.

We use two versions of this galaxy, each with a central BH added. Galaxy 1 includes a $5 \times 10^{8} \, \rm M_{\odot}$ BH. This BH mass is artificially high relative to the value expected for the stellar mass. However, it is representative of the BH masses we expect to see in larger halos and what is associated with quenched galaxies. We choose this high mass to test the scheme under more extreme conditions. Further, it ensures our runs begin in a high accretion/feedback state, making it easier to measure the feedback's effects. Galaxy 2 includes a less massive, $3 \times 10^{7} \, \rm M_{\odot}$ BH. This is more representative of the predicted BH mass for the m12 galaxy \citep[estimated from scaling relations, e.g.][]{Mancini12, Davis19}.

\section{Results}
\label{sec:Results}
In this section, we apply \textsc{ArkBH} to twelve idealized galaxy runs to test the model and to begin investigating its  parameter space (Table~\ref{tab:RunParameters}). We first (\autoref{subsec:Fiducial}) discuss the fiducial run, Fid1, which demonstrates the method's ability to regulate galaxy scale inflows and suppress star formation in our isolated galaxy. With that run we show how \textsc{ArkBH} suppresses star formation through preventative feedback within this run. We then (\autoref{subsec:JetParams}) vary the astrophysical parameters, $v_{\text{w}}$ and $\epsilon_{\text{f}}$, and again explore the results. Finally, we discuss (\autoref{subsec:NumericalParams}) the result of changing the numerical parameters: the recoupling distance or density and the mass resolution factor, $\gamma_{\rm m}$. 

We also carry out three runs without \textsc{ArkBH} feedback for comparison. This includes two runs without any feedback (neither BH nor stellar driven), one in Galaxy 1 and a second in Galaxy 2. These are discussed in \autoref{subsec:Fiducial} and \autoref{subsec:3e7BH}, respectively. We also include a run in Galaxy 1 using the Illustris-TNG BH feedback method (again without stellar feedback), referred to as `TNG', to compare against the fiducial \textsc{ArkBH} run.

\subsection{Fiducial Run}
\label{subsec:Fiducial}

\begin{figure*}
    \centering
    \includegraphics[width=.96\textwidth]{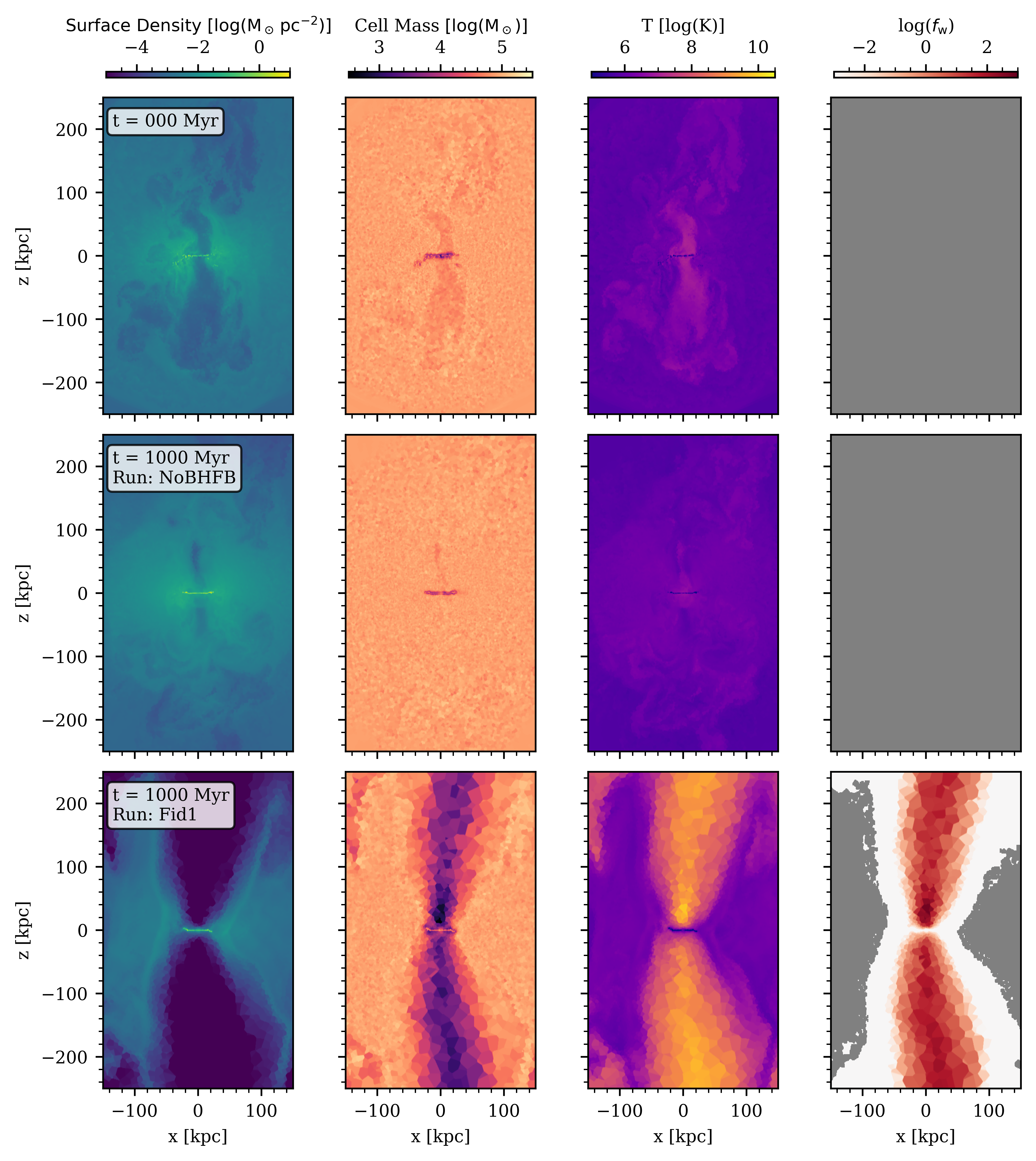}
    \caption{From left to right: Slices taken along the galaxy's x-z plane of the gas surface density, cell mass, gas temperature, and hot wind tracer. The top panel displays these slices at t = 0 Myr. The middle panel displays them again at 1000 Myr in the NoBHFB run, which has feedback turned off. The bottom panel displays them at 1000 Myr in the Fid1 run (see Table~\ref{tab:RunParameters}), in which the Arkenstone BH feedback is added. The wind tracer profile is empty in the top row because no wind particles have launched at t = 0 Myr. Likewise, it is still empty in the middle panel since no wind particles are launched in the NoBHFB run.}
    \label{fig:FidArkImages}
\end{figure*}

\begin{figure}
    \centering
    \includegraphics[width=.48\textwidth]{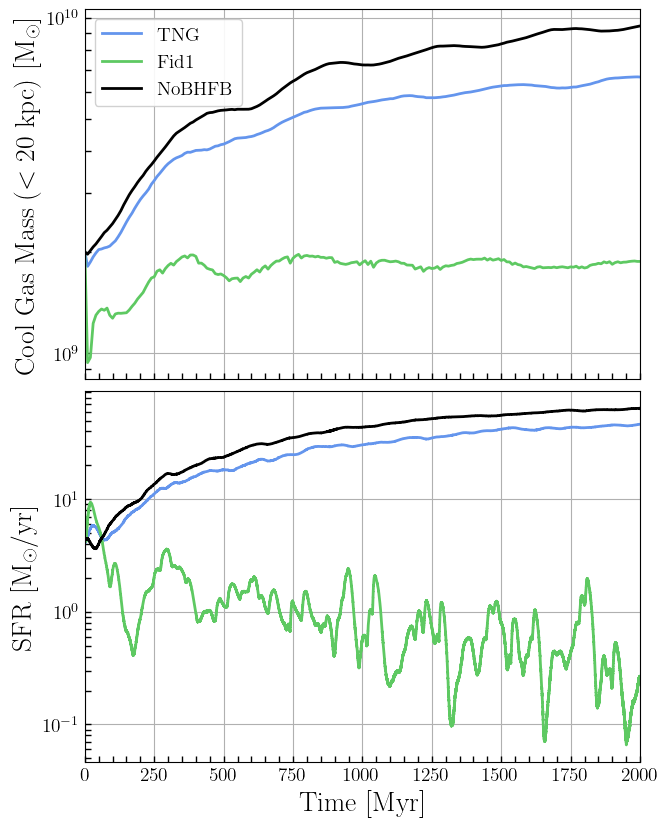}
    \caption{The top panel displays the total cool ($T < 3 \times 10^4$ K) gas mass within 20 kpc of the galaxy center. The bottom panel displays the galaxy's SFR. The Fid1 run (green) has a significantly lower cool gas mass and SFR than the TNG (blue) and NoBHFB (black) runs.}
    \label{fig:FiducialSFR}
    
\end{figure}

\begin{figure}
    \centering
    \includegraphics[width=.48\textwidth]{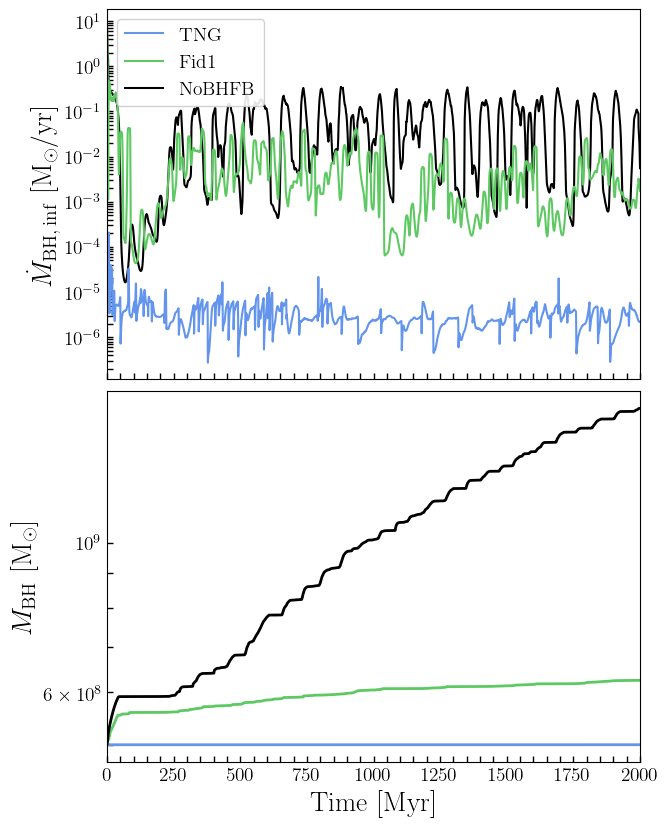}
    \caption{The top panel displays the calculated large scale (Bondi) accretion rate onto the BH, $\dot{M}_\text{BH, inf}$. This is different from the mass actually accreted onto the BH due to mass outflow associated with feedback (see Eq.~\ref{eq:MassFluxes}, Sec.~\ref{subsubsec:ArkBH}). The bottom panel displays the mass of the BH. The black, green, and blue lines again respectively display the NoBHFB, Fid1, and TNG runs. The NoBHFB run has the largest BH growth, followed by the Fid1 and then TNG run. }
    \label{fig:FiducialMassAccretion}
    
\end{figure}

\begin{figure*}
    \centering
    \includegraphics[width=.96\textwidth]{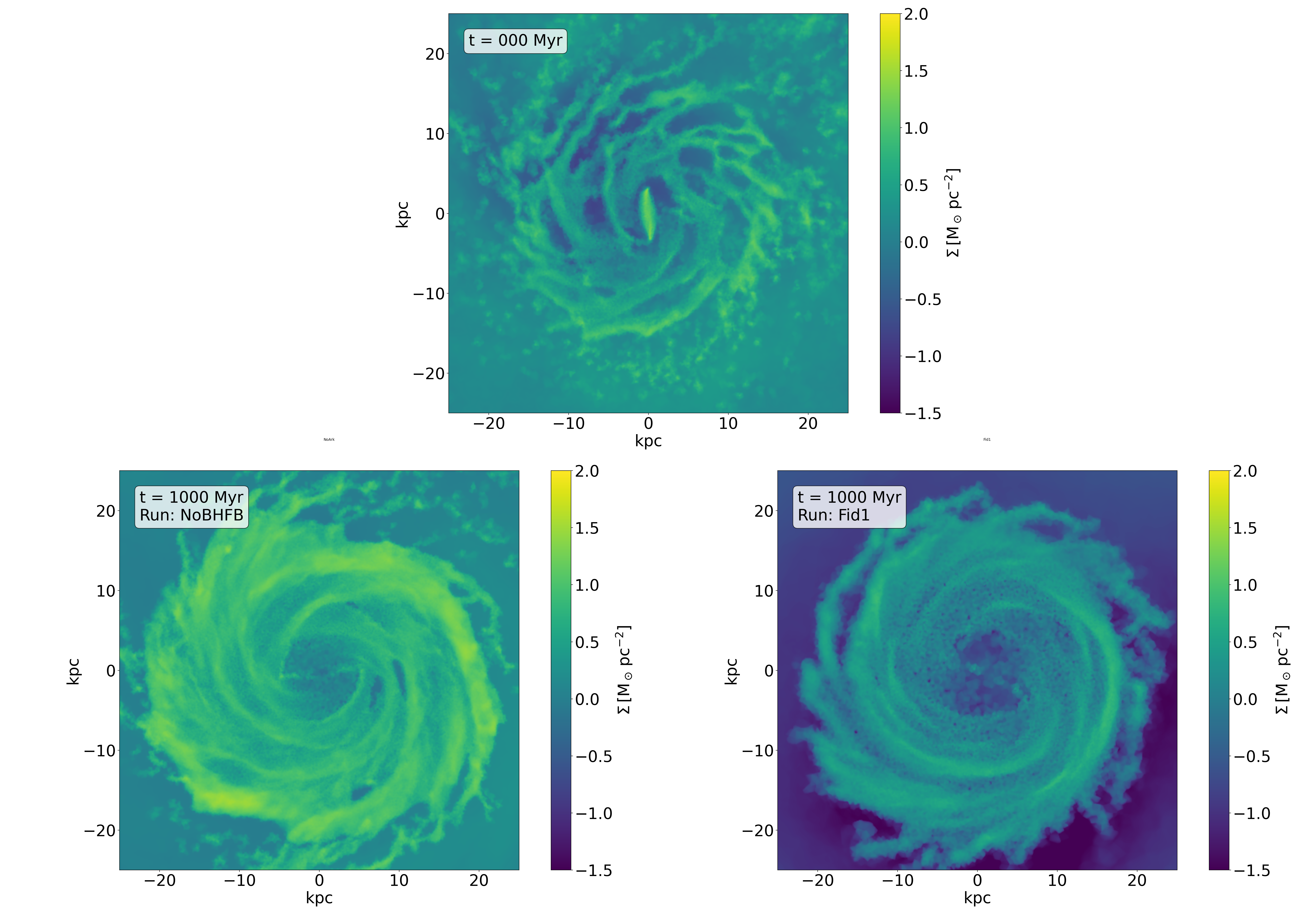}
    \caption{Face-on projections of the gas surface density at t = 0 Myr (top) and then t = 1000 Myr in the NoBHFB (bottom left) and Fid1 (bottom right) runs.}
    \label{fig:NoArkDensity}
\end{figure*}

\begin{figure*}
    \centering
    \includegraphics[width=1.0\textwidth]{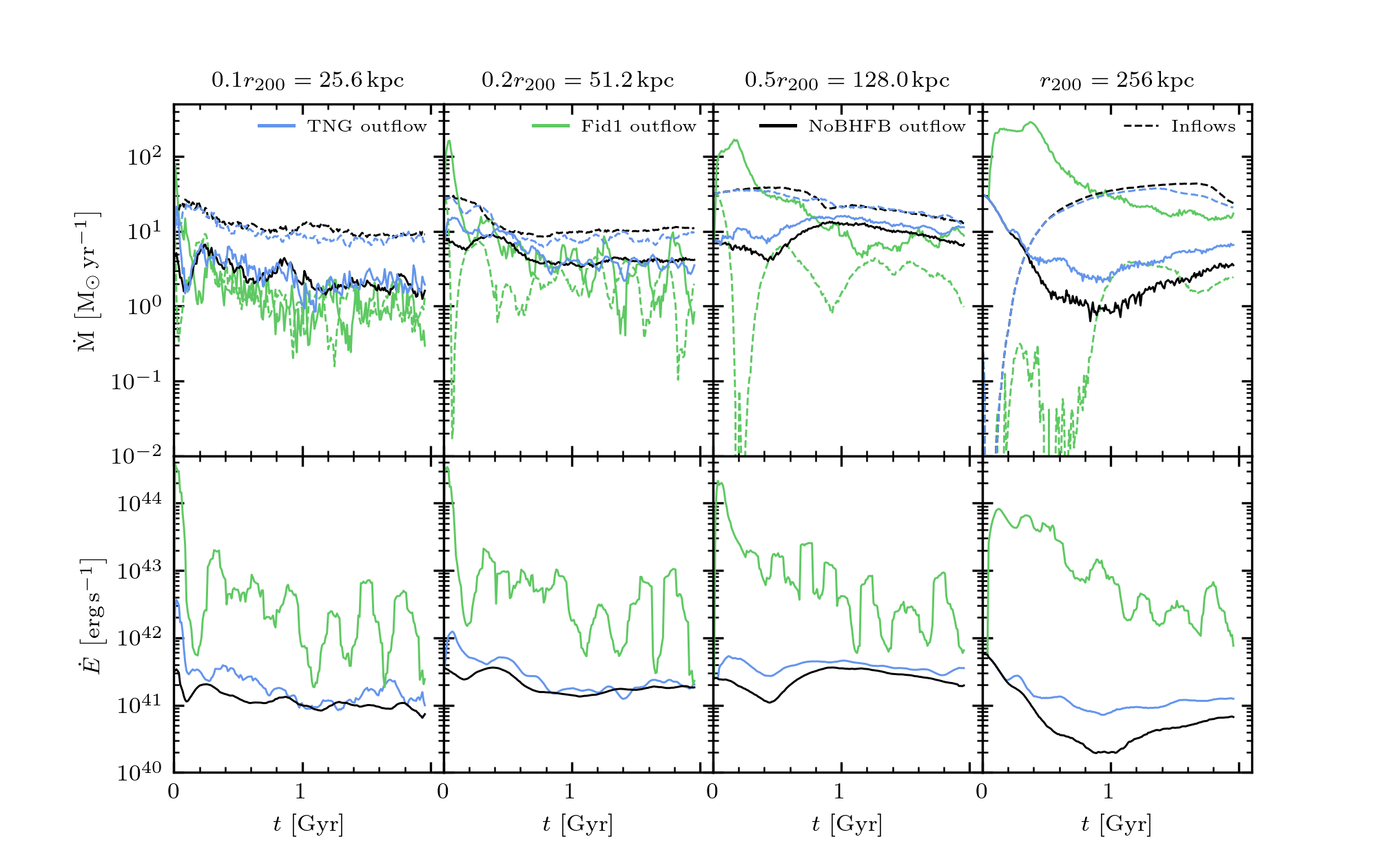}
    \caption{The top row displays the mass inflow (dotted lines) and outflow (solid lines) rates across spherical shells at several radii. The bottom row displays the energy flux across these shells. The black, green, and blue lines display the NoBHFB, Fid1, and TNG runs respectively.}
    \label{fig:FiducialInflowsOutflows}
    
\end{figure*}

\begin{figure*}
    \centering
    \includegraphics[width=.98\textwidth]{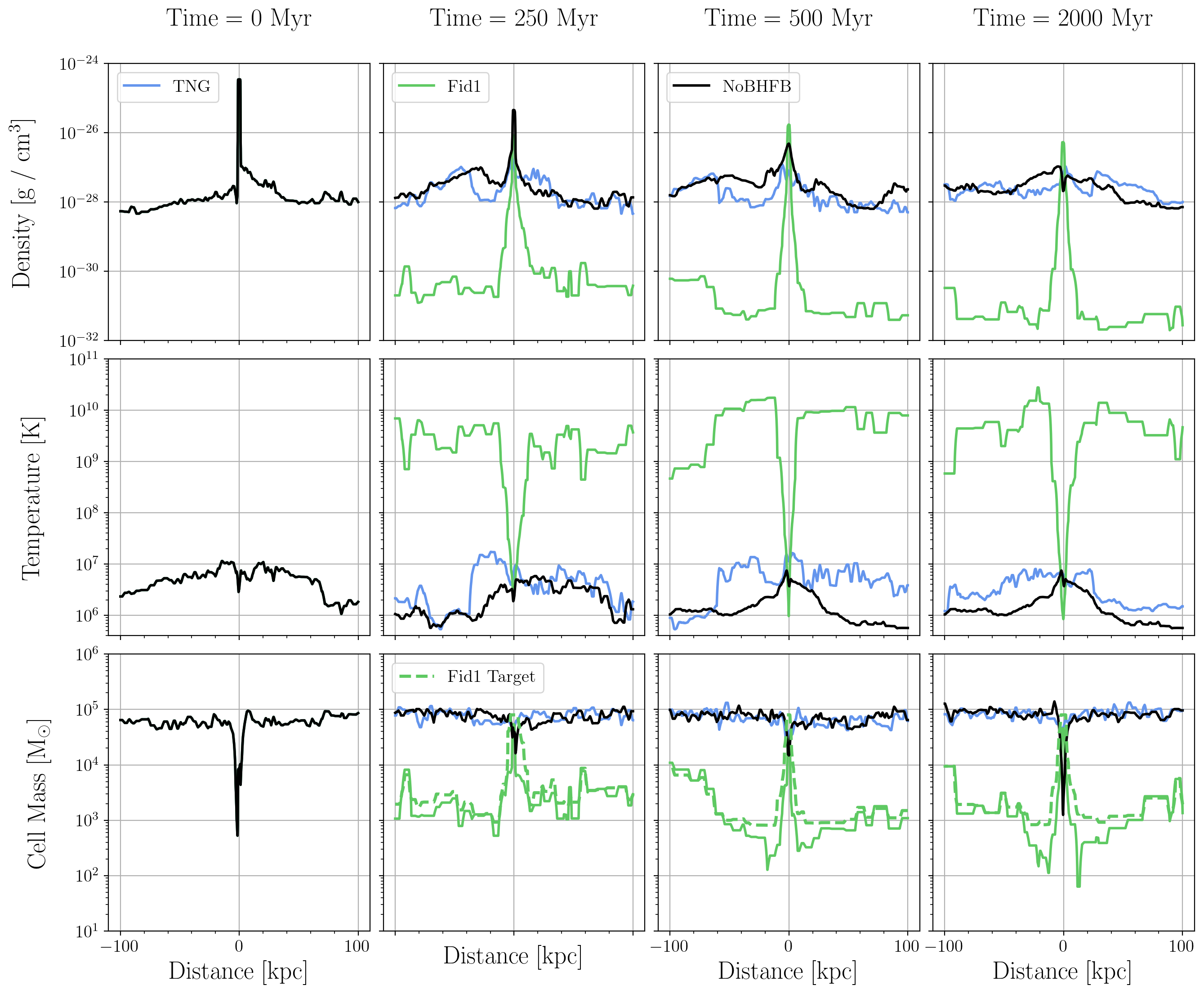}
    \caption{Gas density, temperature, and cell mass profiles for the TNG, Fid1, and NoBHFB runs. In the cell mass profile, we plot both the measured cell mass (solid line) and the value targeted by Arkenstone (dashed line, calculated with eq.~\ref{eq:MassRes}) These profiles are taken along a ray passing through the center of the galaxy and normal to the disk plane. The \textsc{ArkBH} feedback decreases the gas density and increases the gas temperature in regions with hot wind material. The cell masses in these hot wind regions are also lower relative to the NoBHFB and TNG runs.}
    \label{fig:FiducialRays}
    
\end{figure*}

We run a simulation of Galaxy 1 with our fiducial set of parameters (see Fid1, Table~\ref{tab:RunParameters}) for 2~Gyr. This run is designed to demonstrate the effects of a high specific energy jet on the host galaxy. This is compared against two runs---the NoBHFB run excluding both BH and stellar feedback, and the TNG run including only TNG BH feedback. Figure~\ref{fig:FidArkImages} displays the outflow shape and the large impact it has on the gas surrounding the galactic disc. Comparing the Fid1 and NoBHFB runs, \textsc{ArkBH} opens up a significant low density region above and below the disk, while also increasing the gas temperature. These regions broadly trace the expected outflow morphology, though the outflow's impact extends over a larger volume. The side-on slices also display evidence of the stellar wind feedback included in the construction of our ICs. This may impact some of the early evolution as the system readjusts to the current set of physics. However, the effect of this prior feedback is significantly smaller than the observed impact of the BH feedback by $t = 1000 \, \rm Myr$. We investigate the global effect of the \textsc{ArkBH} feedback on the SFR, before focusing on the BH evolution. We then use the mass inflow/outflow rates and gas profiles to discuss how \textsc{ArkBH} suppresses star formation.   

\subsubsection{Effects on Star Formation}
\label{subsec:FidSFREffect}
\textsc{ArkBH} reduces the cool gas available for star formation in our test galaxy (Fig.~\ref{fig:FiducialSFR}). We define this as gas below $3\times10^4 \, \rm K$ within 20 kpc from the galaxy center. This cool gas mass in Fid1 rapidly decreases at the start of the run and then remains well below that of the NoBHFB run. The final cool gas mass in Fid1 is $2.1 \times 10^9 \, \rm M_{\odot}$, as compared to $9.5 \times 10^9 \, \rm M_{\odot}$ in the NoBHFB run. The cool gas mass in our TNG run also decreases relative to the NoBHFB run. However, this difference is much smaller than Fid1's. 

\textsc{ArkBH} also successfully reduces the SFR (Fig.~\ref{fig:FiducialSFR}). In the NoBHFB run, the SFR initially decreases from $4.4 \, \rm M_{\odot}\,yr^{-1}$ to $3.7 \, \rm M_{\odot}yr^{-1}$. It then steadily increases up to $\sim61\, \rm M_{\odot}\,yr^{-1}$ by 2~Gyr as gas inflows fuel further star formation. This increases the total (newly created since $t=0$) stellar mass from $8.3 \times 10^8 \, \rm M_{\odot}$ to $5.3 \times 10^9 \, \rm M_{\odot}$ by 2~Gyr. In contrast, the \textsc{ArkBH} feedback produces an initial small burst of star formation followed by an extended decrease over the rest of the simulation. The final SFR and (new) stellar mass are $3.0 \times 10^{-1} \, \rm M_{\odot}\, yr^{-1}$  and  $10^9 \, \rm M_{\odot}$ respectively, both significantly lower than that of the NoBHFB run.

The SFR in the TNG run decreases slightly relative to the NoBHFB run but remains well above the Fid1 SFR (Fig.~\ref{fig:FiducialSFR}). This is consistent with its relatively high cool gas mass. We discuss further in Appendix~\ref{sec:Appendix} why we do not observe a large decrease in the SFR in the TNG run.

\subsubsection{Black Hole Evolution}
\label{subsubsec:FidBlackHoles}
As mentioned in Section~\ref{subsec:ArkRecoupling}, \textsc{ArkBH} avoids directly modeling the influence of AGN feedback on the ISM. We instead focus on the impact of high specific energy feedback within the CGM (see Section~\ref{sec:intro}). Significant changes to the BH accretion rate\footnote{We reference the BH accretion rate throughout section~\ref{sec:Results}. This refers to the Bondi accretion rate, $\dot{M}_\text{BH, inf}$, unless otherwise mentioned.} (particularly soon after wind particles are launched) signify the feedback is directly impacting the ISM and interacting with gas close to the BH. We therefore expect to produce an initial BH accretion rate in Fid1 close to that of the NoBHFB run. Preventative feedback will eventually decrease the gas inflow into the ISM and BH accretion disk, leading to a divergence between the accretion rates at later times.   

Turning to the BH accretion, we see in Fig.~\ref{fig:FiducialMassAccretion} that without \textsc{ArkBH}, the BH accretion rate initially drops from nearly $10 \rm \, M_{\odot} \, yr^{-1}$ down to $\sim 4\times10^{-1} \rm \, M_{\odot} \, yr^{-1}$ by 10~Myr. It slowly declines for about 45~Myr, after which it begins to fall off more rapidly. The accretion remains low for about 200~Myr, not surpassing $10^{-4} \, \rm M_{\odot}\,yr^{-1}$. This initial drop in accretion is due to the BH depleting its surrounding of gas via accretion at the start of the simulation. The flow of gas back into the galaxy center later balances the accretion rate, leading to higher and more stable accretion for $t \gtrsim 250 \rm \, Myr$. From this point on, the accretion rate fluctuates between $\sim 10^{-4}-1 \, \rm M_{\odot}\,yr^{-1}$. Overall, this results in the BH mass steadily growing and exceeding $1.6 \times 10^9 \, \rm M_{\odot}$ by 2~Gyr (Fig.~\ref{fig:FiducialMassAccretion}). 

The Fid1 and NoBHFB accretion rates are generally very similar until $\sim 250 \, \rm Myr$. After $t=250$ Myr, though both accretion rates vary significantly with time, the Fid1 run has a lower average accretion rate. The Fid1 BH grows $\sim15$ times slower than the NoBHFB run's over the last $1750 \, \rm Myr$ due to both the lower accretion rate in Fid1 and the continued launching of mass through feedback. More importantly however, we demonstrate that we do not actively shut off accretion onto the BH. This is an indication that the feedback is impacting the CGM rather than the ISM. While there may be some fallback effects onto the edge of the ISM from feedback recoupling nearby, we avoid disrupting the central disk with our high specific energy particles.

We confirm this using face-on projections of the gas surface density (Fig.~\ref{fig:NoArkDensity}). While the Fid1 run lowers the average density of the disc and its surroundings relative to NoBHFB, the overall spiral structure remains intact. This preserves a relatively high BH accretion rate which is necessary for the feedback to continue. We also demonstrate that the decrease in the SFR is not simply due to the disk being destroyed. We discuss the mechanisms for suppressing star formation in Section~\ref{subsubsec:QuenchingAnalysis}.

In the TNG run, the average accretion rate is much lower across the entire run. The final BH mass is therefore much smaller than in the other two runs. It only accretes $3.2\times 10^6 \, \rm M_\odot$ in 2~Gyr (Fig.~\ref{fig:FiducialMassAccretion}). This is a result of the TNG kinetic feedback immediately opening a hole in the center of the disk. This significantly reduces the black hole accretion and puts the system in an equilibrium state where feedback prevents the central disk from reforming and increasing the BH accretion rate (again, we discuss this run further in Appendix~\ref{sec:Appendix}).
 
\subsubsection{Regulation of Gas Inflows and Outflows}
\label{subsubsec:QuenchingAnalysis}

We now begin to look at how \textsc{ArkBH} reduces the galaxy's star formation. We compare the mass inflow and outflow rates across several spherical shells in the galaxy, selecting values of 0.1, 0.2, 0.5 and 1.0 times the virial radius (Fig.~\ref{fig:FiducialInflowsOutflows}). In the NoBHFB run, the mass inflow is consistently above the mass outflow rate across the first three shells. In the outermost shell, we see that the mass outflow is initially larger for $\sim 300 \, \rm Myr$. This is a transient phase produced by our initial conditions setup. As mentioned in Section~\ref{subsec:TestGalaxies}, stellar wind feedback was included when generating the initial conditions but is then turned off in our runs. The mass inflow rate rises quickly in its absence, growing larger than the outflow rate. This is consistent with the earlier SFR and BH results. We see that, in the absence of feedback, the CGM feeds gas back into the ISM and onto the BH. 

\textsc{ArkBH} significantly changes the balance of in- and out-flowing material in the galaxy. Across all shells, there is a clear spike in the mass outflow rate and drop in the inflow rate when the run begins. In addition, the mass outflow rate grows with radius. This is due to the jet feedback, which adds large amounts of energy to the gas, lifting up the material, carrying it outwards. The mass inflow rate drops accordingly. This is a clear display of the preventative feedback we are modeling since the jet itself carries little mass. This is backed up by the energy outflow rates, which are significantly larger than in the NoBHFB run across each shell. This energy injection significantly reduces the mass inflow, preventing gas from accreting onto the disk. This drives the decrease in star formation described earlier. We also compare the amount of mass ejected as wind particles (i.e. direct BH outflows), $\sim3.6 \times 10^6 \, \rm M_\odot$, to the difference in final stellar mass between the Fid1 and NoBHFB runs. The final stellar mass is $4.3 \times 10^9 \, \rm M_\odot$ lower in the Fid1 run, demonstrating that the difference in star formation cannot simply be explained by the initial injection of jet material. 

We compare the effects on the gas density and temperature, as well as the cell mass (Fig.~\ref{fig:FiducialRays}). These are evaluated along narrow rays passing through the galaxy center, perpendicular to the plane of the disk (aligned with the jet direction). In the Fid1 run, the gas density along the entire ray decreases out to $r \simeq 25 \, \rm kpc$ in either direction. The temperature has a corresponding increase, similarly rising to $\sim 10^{9-10} \, \rm K$ near $25 \, \rm kpc$. This heating effect extends more generally into the surrounding regions and decreases the cool gas mass within the ISM (see also Sec.~\ref{subsec:FidSFREffect}).



We also see that the cell masses along the ray in Fid1 are significantly smaller than in the NoBHFB run. Further, they closely follow the cell mass dictated by the cell's hot wind tracer (green dashed lines in bottom row of Fig.~\ref{fig:FiducialRays}). This demonstrates that the Arkenstone method is correctly refining the cell mass based on the hot wind tracer.

\subsection{Jet Parameters}
\label{subsec:JetParams}

\begin{figure}
    \centering
    \includegraphics[width=.48\textwidth]{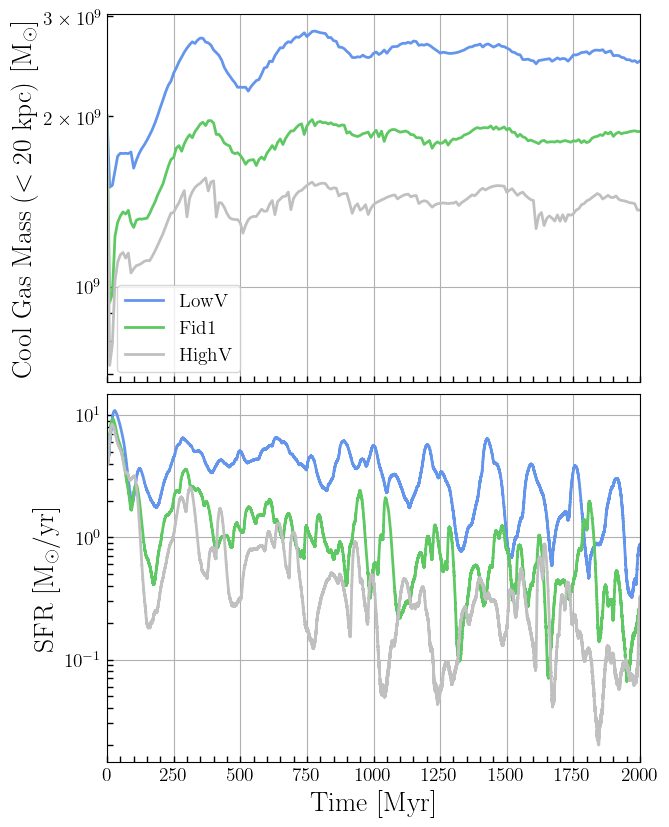}
    \caption{The cool gas mass within 20 kpc of the galaxy center (top panel) and the galaxy's SFR (bottom panel) in the wind velocity test runs. The blue, green, and gray lines display the LowV, Fid1, and HighV runs.}
    \label{fig:VelocitiesSFR}
    
\end{figure}

\begin{figure}
    \centering
    \includegraphics[width=.48\textwidth]{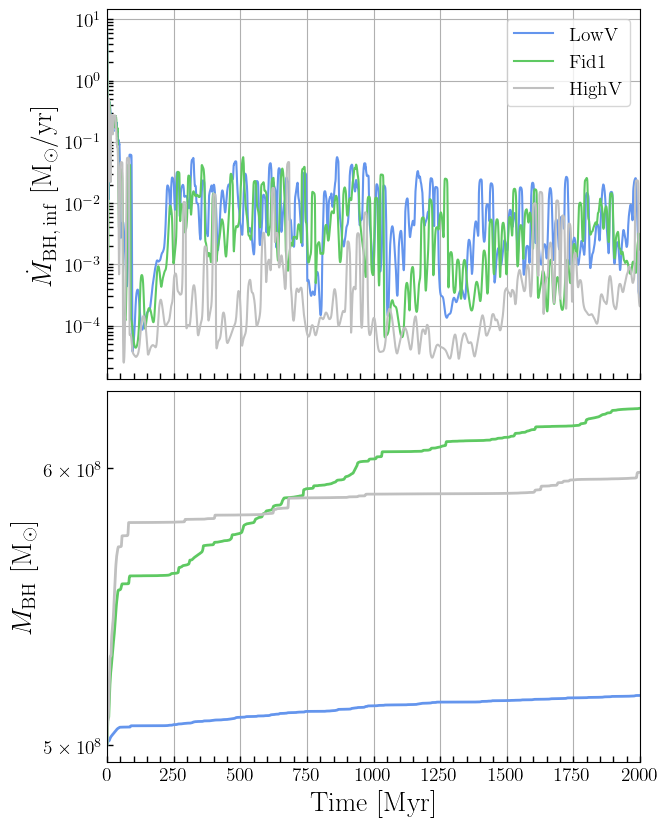}
    \caption{The top and bottom panels display the BH's Bondi accretion rate ($\dot{M}_\text{BH, inf}$) and total mass, as in Fig.~\ref{fig:FiducialMassAccretion}. The blue, green, and gray lines again display the LowV, Fid1, and HighV runs,  as in Fig.~\ref{fig:VelocitiesSFR}.}
    \label{fig:VelocitiesMassAccretion}
    
\end{figure}

    

\begin{figure}
    \centering
    \includegraphics[width=.48\textwidth]{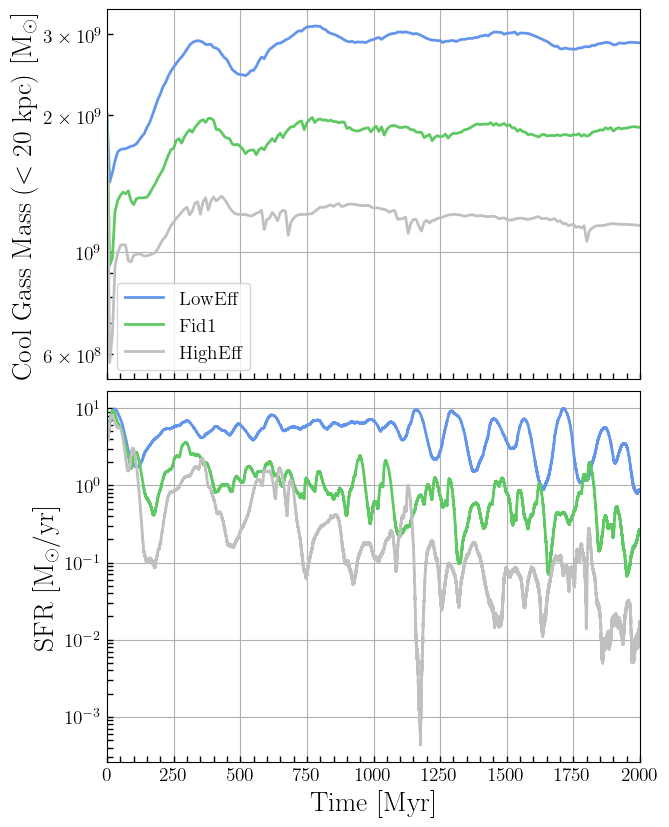}
    \caption{The cool gas mass within 20 kpc of the galaxy center (top panel) and the galaxy's SFR (bottom panel) in the feedback efficiency test runs. The blue, green, and gray lines display the LowEff, Fid1, and HighEff runs respectively.}
    \label{fig:EfficienciesSFR}
    
\end{figure}

\begin{figure}
    \centering
    \includegraphics[width=.48\textwidth]{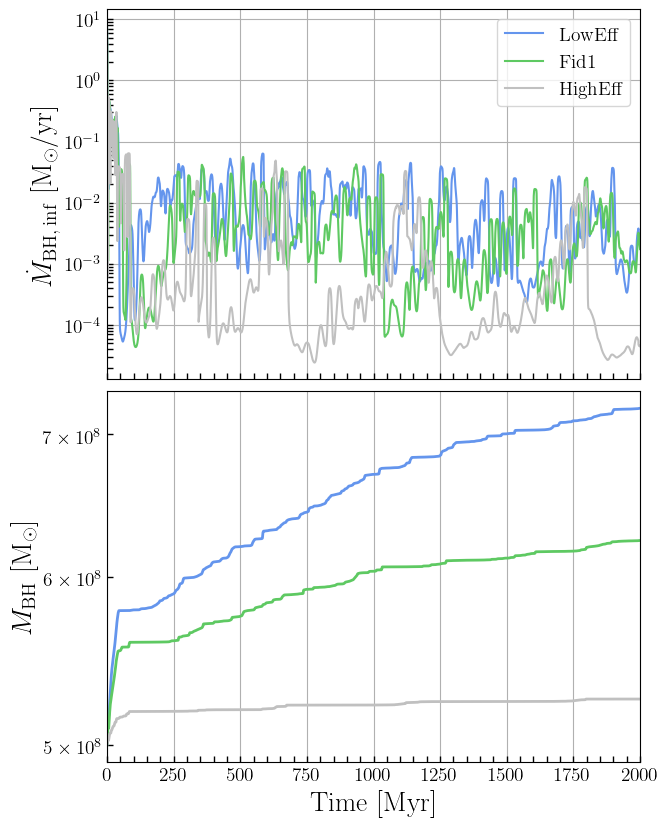}
    \caption{The top and bottom panels display the BH's Bondi accretion rate and total mass, as in Fig.~\ref{fig:FiducialMassAccretion}. The blue, green, and gray lines display the LowEff, Fid1, and HighEff runs respectively, as in Fig.~\ref{fig:EfficienciesSFR}.}
    \label{fig:EfficienciesMassAccretion}

\end{figure}

    

We next vary the wind particle velocity ($v_{\text{w}}$) and outflow efficiency ($\epsilon_{\text{f}}$) to explore how these parameters affect our results. 

\subsubsection{Wind Particle Velocity}
\label{subsubsec:WindVelocity}

We first compare the different wind velocity runs. This includes wind velocities of $10^4 \rm \, km\,s^{-1}$ (LowV), $3\times10^4 \rm \, km\,s^{-1}$ (Fid1), and $10^5 \rm \, km\,s^{-1}$ (HighV), as listed in Table~\ref{tab:RunParameters}. We keep the efficiency $\epsilon_{\text{f}}$ fixed; however, note that higher wind velocities correspond to both higher specific energies and higher total energies in terms of the Bondi accretion rate (Eq.~\ref{eq:EdotkBondi}).

Increasing the velocity further suppresses star formation and reduces the cool gas mass (Fig.~\ref{fig:VelocitiesSFR}). Further, the shape of the trends in cool gas mass appears the same, albeit with a vertical shift (decrease) for higher velocity winds. The SFRs exhibit slightly more variation across the three runs but there is still the same general downwards trend. The three runs also begin to converge more at late times, with the Fid1 run in particular settling at a similar star formation to the HighV run. This is expected, since Fid1 will continue to launch more high specific energy outflows than the HighV run due to its larger BH accretion rate (discussed below). These two runs both exhibit lower SFRs than the LowV run. However, there is clear suppression of star formation across this entire parameter space, including in the LowV run. 

Early on ($t\lesssim 150 \, \rm Myr$), the three runs have very similar BH accretion rates, with the Fid1 and HighV runs exhibiting a slightly `burstier' accretion rate (Fig.~\ref{fig:VelocitiesMassAccretion}). However, the HighV run's accretion rate diverges from that of the other two runs past this point. It remains lower (at $\simeq 10^{-(3-4)} \, \rm M_{\odot}\, yr^{-1}$) than the other two runs for most of the remaining time, excepting several spikes of comparable accretion ($\geq 10^{-2} \, \rm 
M_{\odot}\, yr^{-1}$). 

The BH mass therefore initially grows fastest in the HighV run, followed by the Fid1 and LowV runs. This is a consequence of the jet parameterization. For a fixed $\epsilon_f$, decreasing the wind particle velocity produces a lower specific energy outflow. More mass must be ejected to produce the same $\dot{E}$. To expand on this, for a fixed large scale accretion rate (i.e. Bondi rate), $\dot{M}_{\text{BH, inf}}$, the effective accretion rate, $\dot{M}_{\text{BH}}$, is proportional to $1/(1+\psi)$. Because $\psi$ is proportional to $v_\text{w}^{-2}$, decreasing the wind velocity decreases the effective accretion rate. Since the runs initially have similar accretion rates, more mass is initially ejected in feedback (i.e. larger $\dot{M}_\text{wind}$). However, the Bondi accretion rate in HighV then drops relative to the other two runs, allowing the Fid1 BH mass to exceed the HighV run's later on.

We again measure the gas density, temperature, and cell mass along rays passing through the center of the galaxy, perpendicular to the disk. This shows similar trends to those described in Section~\ref{subsubsec:QuenchingAnalysis}, so we do not show it for brevity. The HighV run exhibits a smaller initial decrease in the gas density and a smaller increase in temperature than the Fid1 and LowV runs. By 2~Gyr, its ray profiles are very similar to those of the lower velocity runs. Across the three runs, the lower velocity runs generally display more stable gas profiles. Splitting up the feedback into more particles, each with less kinetic energy produces a less volatile or `bursty' behavior. The outflow region in the LowV run also maintains a more constant hot wind tracer value and therefore mass resolution over the course of the simulation. 

We demonstrate the ability to suppress star formation across the range of wind velocities. However, we also show that this has an effect on other galaxy properties such as the BH growth and gas profiles. For the former, the high velocity feedback shows the ability of the BH to self-regulate its growth through feedback. It alternates periods of low accretion with short bursts of energetic feedback. The BH mass and accretion rate can thus provide useful constraints on the choice of wind velocity in future simulations.   

\subsubsection{Feedback Efficiency}
\label{subsubsec:OutflowEfficiency}
 
Increasing the feedback efficiency, $\epsilon_f$, has the expected effect on the cool gas mass and SFR (Fig.~\ref{fig:EfficienciesSFR})---it decreases the available cool gas mass over the simulated time. This leads to a further decrease in the SFR relative to the NoBHFB run. In all three runs (LowEff, Fid1, and HighEff), the SFR is continuing to decrease by 2~Gyr, demonstrating the outflows are continuing to suppress star formation over a longer timescale.

We note that there is a more significant decrease in the star formation here in the HighEff run as compared to the HighV run. This is because the associated outflow energy and mass fluxes (Eqns.\ref{eq:MassFluxes}, \ref{eq:EdotthBondi}) are larger in the HighEff run. This drives a larger decrease in the mass inflow and also the cool gas mass. Comparing the BH accretion rates, the HighEff run displays more bursts of accretion, which then self-regulates due to the feedback. 

Increasing the efficiency generally decreases BH growth and leads to more strongly self-regulated accretion. In the HighEff run, the accretion rate exhibits a more bursty pattern (Fig.~\ref{fig:EfficienciesMassAccretion}), with short lived spikes followed by more extended periods of low accretion. This reflects the cycle of stronger feedback bursts suppressing accretion for larger periods of time. On average however, the accretion is clearly lowest in the HighEff run due to these extended periods of lower accretion.

The differences in BH mass follow directly from these results. The higher efficiency causes more material to be ejected (for the same large-scale accretion rate, $\dot{M}_\text{BH, inf}$), and thus less mass is actually accreted onto the BH. The BH mass is therefore largest in the LowEff run and lowest in the HighEff run. 

The effect on the gas above and below the disk is similar to that in the three velocity runs, so we again omit the plot for brevity. The LowEff run has the smallest effect on the gas properties. The outflows in this run have both lower mass fluxes for a fixed Bondi accretion rate (Eq.~\ref{eq:MassFluxes}) and also a lower energy budget for that feedback (Eqs.\ref{eq:Edotka},~\ref{eq:Edotth}). The Fid1 and HighEff runs produce relatively similar results due to the larger self-regulation of accretion and feedback that occurs in the HighEff run.

\subsection{Numerical Parameters}
\label{subsec:NumericalParams}

\begin{figure}
    \centering
    \includegraphics[width=.48\textwidth]{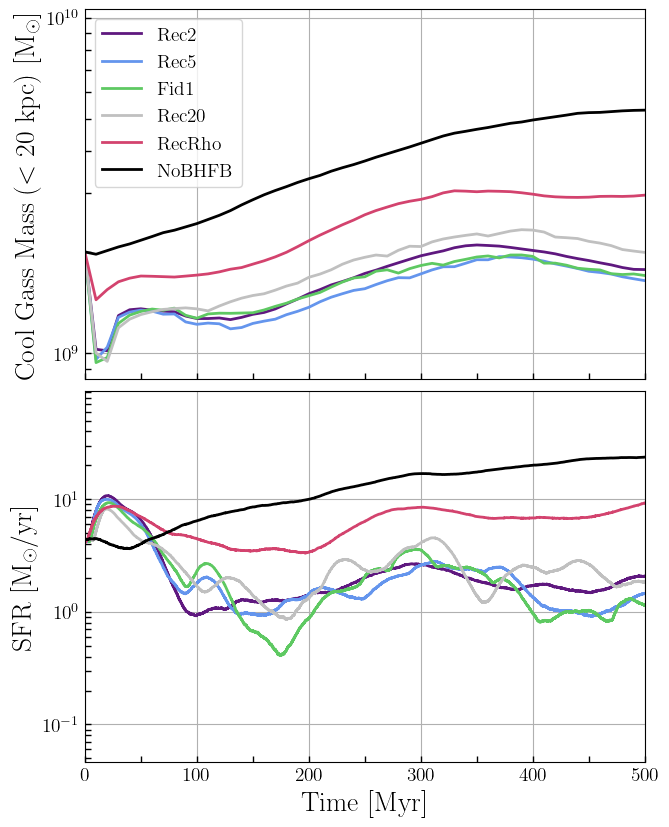}
    \caption{The cool gas mass within 20 kpc of the galaxy center (top panel) and the galaxy's SFR (bottom panel) in the recoupling runs. The four radial recoupling runs (Rec2, Rec5, Fid1 -- 10 kpc, and Rec20) and the density recoupling run (RecRho) are shown. The NoBHFB run is also plotted for comparison.}
    \label{fig:RecouplingSFR}
    
\end{figure}

\begin{figure}
    \centering
    \includegraphics[width=.48\textwidth]{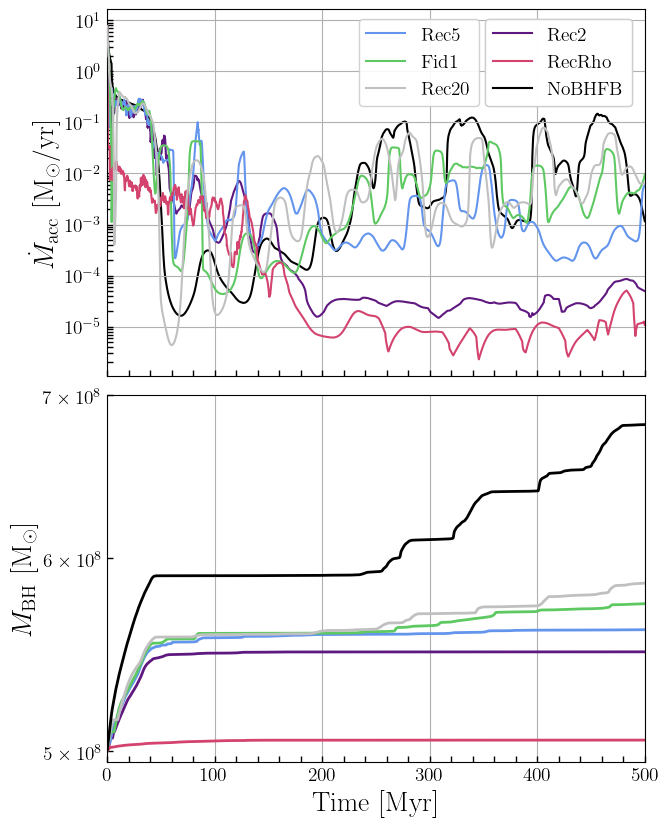}
    \caption{The top and bottom panels display the BH's Bondi accretion rate and total mass, as in Fig.~\ref{fig:FiducialMassAccretion}. The four radial recoupling runs (Rec2, Rec5, Fid1 -- 10 kpc, and Rec20), the density recoupling run (RecRho), and the NoBHFB run are again shown.}
    \label{fig:RecouplingMassAccretion}
    
\end{figure}

Lastly, we explore the numerical parameter space. This is defined as the recoupling criteria ($\text{R}_{\text{rec}}$ or $\rho_{\text{rec}}$) and the mass factor ($\gamma_{\text{m}}$). These runs (excluding Fid1) span 500~Myr as opposed to 2~Gyr. We find that changing the mass factor produces little variation in our results for $\gamma_\mathrm{m} = 0.001$–$0.1$, while the $\gamma_\mathrm{m} = 1.0$ case shows more significant deviations. We discuss this further in Appendix~\ref{subsubsec:Appendix2}.

\subsubsection{Recoupling Distance}
\label{subsubsec: RecouplingDistance}

We first vary the recoupling distance, testing $\text{R}_{\text{rec}}$ = 2, 5, 10, and 20 kpc. We also add a run using a recoupling density,  $\rho_{\,\text{rec}}$ = 0.01.  These parameters set where the wind/jet particles recouple into the galaxy. Again, the recoupling density translates to wind particles recoupling back into the simulation at a fraction (here, 1/100th) of the `critical density' for star formation, $\rho_\mathrm{crit}$. At $t = 0 , \rm Myr$, this occurs at a radius of $\sim 1.5 \, \rm kpc$, determined by the first gas cell within the biconical launch region to drop below this density threshold.   

The cool gas mass and SFR is lowered by the \textsc{ArkBH} feedback across all of the runs (Fig.~\ref{fig:RecouplingSFR}). This suppression is partly driven by a decrease in gas density, both in the central regions and more broadly throughout the galaxy. The 4 radial recoupling runs in particular exhibit similar trends in their cool gas masses and SFRs. Their SFRs sit on the order of $1\, \rm M_{\odot} \, yr^{-1}$. Noticeably, the Rec20 run has the highest cool gas mass of the 4 recoupling runs, albeit by a small amount. It is logical we see the smallest effect here. The wind particles in this run are recoupling at the border of the region we calculate the cool gas mass within. It is also significant that these wind particles still reduce the cool gas mass, demonstrating that the preventative feedback slows the rate at which cool gas flows into the ISM and adds to the measured cool gas mass. 

The RecRho run has a similar shaped curve for the cool gas mass and SFR. However, it is not able to suppress star formation as effectively as the radial recoupling runs. The cool gas mass and SFR are both relatively high and continuing to increase at the end of the 500~Myr (Fig.~\ref{fig:RecouplingSFR}). This is due to the impact of this recoupling choice on the BH accretion (Fig.~\ref{fig:RecouplingMassAccretion}), which we discuss next.

We first look at the initial BH accretion rate for each run during the time period $0 \le t\le 40 \, \rm Myr$. The BH accretion rate drops the most significantly and soonest in the RecRho run, sitting at $\sim 10^{-(2-3)} \, \rm M_{\odot}\,yr^{-1}$. This is at least an order of magnitude lower than the BH accretion rate in the other runs for the majority of this time period. This is because the wind particles start to recouple within the ISM, reducing the local density around the BH. Subsequent wind particles then continue to recouple inside the disk/ISM rather than within the CGM, which we discuss further below. 

The accretion rates are similar across the radial recoupling runs (Rec2, Rec5, Fid1, and Rec20) for the first $\sim 50 \rm \,Myr$. However, we observe pronounced initial drops in accretion in the Fid1 and Rec20 runs. This is likely due to pressure fronts from particles at smaller recoupling distances pushing material towards the disk and BH more quickly, increasing and sustaining the higher accretion rate. This process takes longer in the Fid1 ($\rm R_{rec} = \,10 kpc$) and Rec20 runs. However, the Fid1 and Rec20 accretion rates quickly rise back towards that of Rec2 and Rec5. 

The accretion rates diverge more after this initial time period. The Fid1 and Rec20 runs again have lower accretion rates until $\sim 200\, \rm Myr$, when they pass the Rec2 and Rec5 runs. The Rec2 run steadily decreases after the initial accretion burst and approaches the RecRho accretion rate near 200 Myr. It remains above RecRho's accretion rate but clearly shows that the feedback has disrupted the disk, and consequently the BH accretion. The other 3 radial recoupling runs appear much more similar across the 500~Myr run, with Rec5 having the lowest final accretion rate of the three.

Holistically, the Fid1 (i.e., 10 kpc recoupling) and Rec20 runs produce BH accretion rates closer to the NoBHFB rate (Fig.~\ref{fig:RecouplingMassAccretion}). However, as described above, both runs exhibit a reduction in their cool gas mass relative to the NoBHFB run (Fig.~\ref{fig:RecouplingSFR}). This highlights that the suppression of cool gas mass is not only driven by changes in BH accretion and therefore the amount of feedback, but also by where the feedback energy is deposited. In these larger recoupling radius runs, energy is deposited further into the CGM, where it can heat gas and prevent inflows of cool gas. As a result, the supply of cool gas to the ISM is reduced even though the BH accretion rate is similar to the run without feedback. 

Out of the four recoupling distance runs, the Rec2 recoupling run diverges the furthest from the NoBHFB BH accretion rate, producing an accretion rate and BH mass significantly lower than the other runs. We find that recoupling at smaller radii allows the high specific energy material to interact more with material in the ISM and local to the black hole. This decreases the gas density surrounding the black hole and produces the decrease in the BH accretion rate and final mass. This then carries over into the feedback, reducing the number of ejected wind particles. 

However, the suppression of accretion and therefore feedback is the most extreme in the RecRho run. In this case, as the central density decreases, wind particles recouple at progressively smaller radii. This produces a runaway effect in which feedback continues to couple increasingly close to the BH. This rapidly evacuates the surrounding gas and drives down the accretion rate and amount of feedback launched.

By contrast, the Rec2 run still suppresses star formation in the galaxy despite the reduced BH accretion rate at later times. This is because unlike in the RecRho run, the fixed 2 kpc recoupling distance in the Rec2 run prevents this inward migration of the coupling location. While feedback still interacts strongly with the ISM and reduces the gas supply near the BH, it does so at a finite (and larger radius than in RecRho). As a result, the Rec2 run sustains a higher BH accretion rate for longer, thus launching more outflow material. The outflow also is injected out of the disk, allowing it to more effectively heat gas in the CGM and suppress the inflow of gas.

These results imply that by recoupling at 5 kpc and beyond, the high specific energy material is exerting more influence in the CGM rather than within the ISM and near the BH. This is in line with our previously stated goal of modeling jet feedback that deposits most of its energy in the CGM (and leaving the ISM interaction to a future subgrid model).

\subsection{Lower Mass Black Hole}
\label{subsec:3e7BH}

\begin{figure}
    \centering
    \includegraphics[width=.48\textwidth]{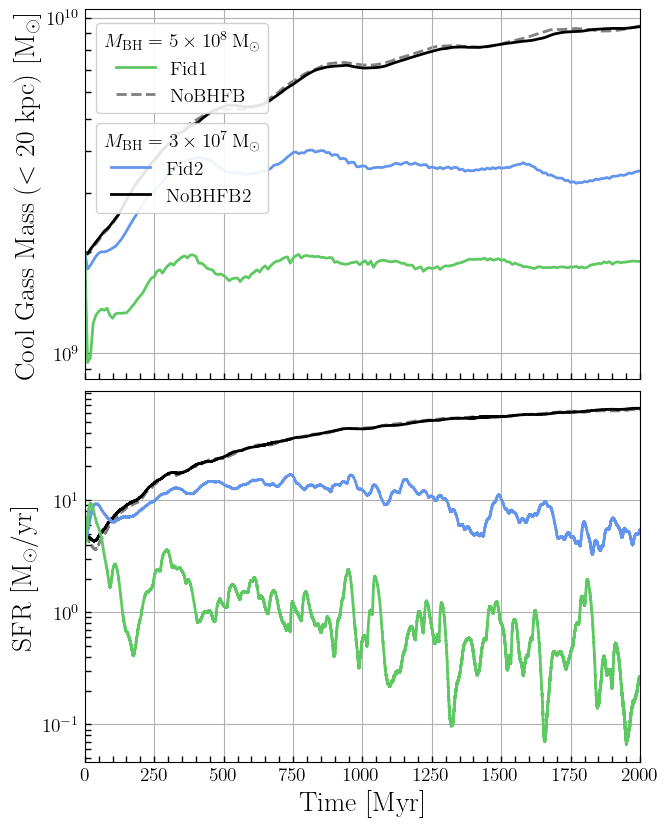}
    \caption{The cool gas mass within 20 kpc of the galaxy center (top panel) and the galaxy's SFR (bottom panel) in the Galaxy 2 runs (NoBHFB2 and Fid2). This corresponds to the NoBHFB and Fid1 runs but in a galaxy with a lower BH mass ($M_\text{BH} = 3\times 10^7 \,\rm M_\odot$). Both sets are included to allow for comparison.}
    \label{fig:3e7SFR}
    
\end{figure}

\begin{figure}
    \centering
    \includegraphics[width=.48\textwidth]{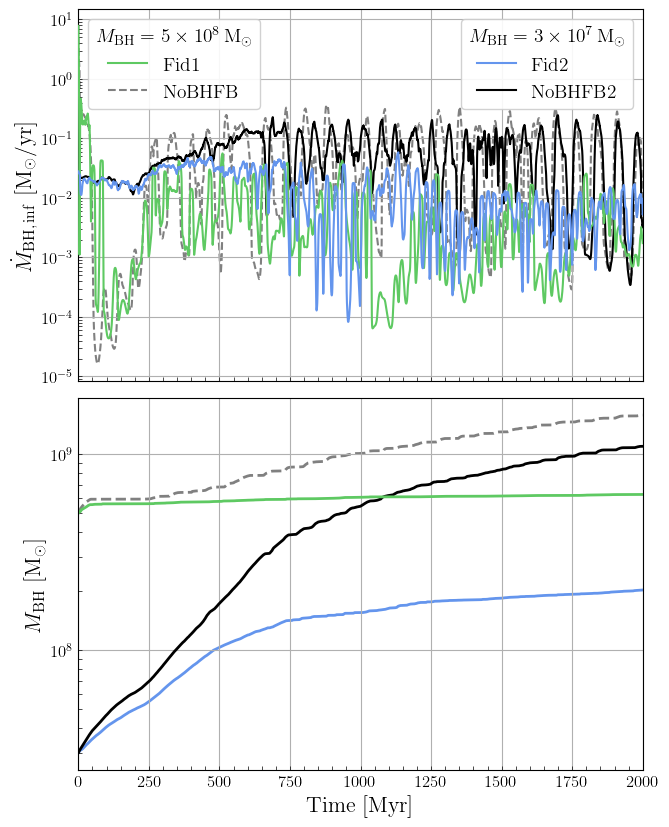}
    \caption{The top and bottom panels display the BH's Bondi accretion rate and total mass. The NoBHFB2 and Fid2 runs ($M_\text{BH} = 3\times 10^7 \,\rm M_\odot$) are plotted alongside the NoBHFB and Fid1 runs to show the effect of the lower initial BH mass.}
    \label{fig:3e7MassAccretion}
    
\end{figure}

Finally, we include two runs to explore the impact of Arkenstone in the same galaxy with a smaller black hole (see Galaxy 2 in Sec.~\ref{subsec:TestGalaxies}). The BH mass, $3\times 10^7 \, \rm M_\odot$, is more representative of the expected BH size for our $10^{12} \, \rm M_\odot$ halo. The first run in this galaxy, Fid2, uses the same parameters as Fid1. The second run, NoBHFB2, has no feedback and uses the same parameters as NoBHFB.   

Again, \textsc{ArkBH} reduces the SFR and the available cool gas mass (Fig.~\ref{fig:3e7SFR}). The final SFR in Fid2 is  $3.4 \, \rm M_\odot \,yr^{-1}$, lower than its initial SFR of $4.4 \, \rm M_\odot \,yr^{-1}$. This is significantly lower than the NoBHFB2 run's final SFR, which increases from the same initial value to $6.5 \times 10^1\, \rm M_\odot \,yr^{-1}$ by 2~Gyr.

The BH accretion rates early on in both the Fid2 and NoBHFB2 simulations are much more stable than in the Galaxy 1 runs due to the lower BH mass (Fig.~\ref{fig:3e7MassAccretion}). In the Bondi accretion framework implemented in \textsc{Arepo}, the accretion rate scales as $\dot{M}_\mathrm{Bondi} \propto M_\mathrm{BH}^2$, making it highly sensitive to BH mass. As a result, the lower mass BH has a reduced initial accretion rate and injects less feedback energy. This leads to less disruption of the surrounding gas and a more stable accretion history. 

The initial accretion rate of $2.7\times10^{-2} \, \rm M_\odot \,yr^{-1}$ is much lower than the $7.8 \, \rm M_\odot \,yr^{-1}$ in the higher mass BH runs. It then slowly increases but remains in the range of $10^{-(1-2)} \, \rm M_\odot \,yr^{-1}$ until $\sim 500 \,\rm Myr$. The Fid2 and NoBHFB2 BH accretion rates remain very similar to this point and then begin to diverge. After this point, the Fid2 accretion rate is lower on average than in NoBHFB2 (but higher than Fid1's). This produces final BH masses of $10^9 \rm \, M_\odot$ and $2\times10^8\,\rm M_\odot$, respectively, in the NoBHFB2 and Fid2 runs.

The \textsc{ArkBH} model has a very similar effect on the gas ray profiles to what we discuss for Galaxy 1 (sec.~\ref{subsubsec:QuenchingAnalysis}). The main difference is less variation in Galaxy 2 between the profiles at different times. The gas temperature remains high and the density low outside of the recoupling radius (along the rays). This is a sign that the high specific energy jet material is successfully heating gas within the outflow region.

\section{Discussion}
\label{sec:Discussion}
In this section, we compare \textsc{ArkBH} against several other subgrid models for BH feedback. We then discuss limitations of this work.

\subsection{Comparison to other Models}
\label{subsec:ModelComparisons}

There are numerous existing AGN feedback models (e.g. Illustris-TNG, SIMBA, EAGLE, Mistral), which vary both in the type(s) of feedback they implement and also in how and where (i.e. locally in ISM or at larger scales) energy and momentum are coupled to the surrounding gas. We use these distinctions to guide our comparisons to \textsc{ArkBH}. For the purposes of comparing to \textsc{ArkBH}, we focus when possible on their prescriptions for kinetic (also sometimes referred to as `mechanical') mode feedback. This is usually associated with low accretion rates, collimated outflows, and high energy jets, which we focus on in this work. Some models however do not include a kinetic feedback mode, so we instead focus on the other relevant pieces. We also describe in more detail the \citet{Choi15} and MISTRAL \citep{Farcy25} models, which share similarities with our numerical framework.

\textsc{ArkBH} uses nearly the same physical framework as \citet{Choi15} and \citet{Farcy25} to associate how much mass and energy is accreted onto the BH versus and how much is `ejected' as feedback. We simply add an additional split between thermal and kinetic energy. Our approach primarily differs in how this energy and momentum is coupled back into the galaxy. Instead of depositing feedback directly into ISM gas cells through a `kick', we create wind particles with a set mass, energy, and momentum. These particles are hydro-decoupled from the simulation until they enter the CGM. This stops the jets from interacting with and entraining a (possibly nonphysically) large amount of ISM material due to the region's low resolution. As stated earlier, our model instead emphasizes how BH jets drive preventative feedback through their interactions with CGM gas.

\textsc{simba}'s BH feedback model uses a kinetic feedback prescription to model both the radiatively efficient (quasar) and inefficient (jet) modes. At Eddington ratios less than 0.2, the feedback transitions to high velocity jets, maxing out at velocities $>7000 \, \rm km \, s^{-1}$ for $f_\text{Edd} \lesssim 0.02$. The outflow temperature is raised to the virial temperature of the halo. This jet mode also is only activated in BHs larger than $10^{7.5} \rm \, M_\odot$.  We again note that our model is capable of launching and tracking outflows with different launch velocities and energies, allowing for a similar dichotomy. We instead focus only on the jet feedback mode in this work. Comparing \textsc{ArkBH} to \textsc{simba} further, we similarly launch feedback in the form of wind particles. We directly set the jet velocity and then use the energy loading factors to adjust the outflow thermal energy. To launch particles, \textsc{simba} randomly selects which particles within the BH kernel to eject. The launch probability is based on the calculated accretion rate and the fraction of mass accreted by the BH. In the jet mode, these particles are ejected (anti)parallel to the disk angular momentum to create a bipolar outflow. There is no opening angle but the outflow has a discrete width due to particles being kicked from their location in the smoothing kernel. By contrast, we assign an opening angle to our jet. This also allows us to vary the opening angle depending on the mode of feedback. \textsc{simba} also applies hydrodynamic and radiative cooling decoupling to its ejected wind particles. At $z=0$, their jets have a maximum recoupling distance of $\sim 11 \, \rm kpc$. This mitigates the previously discussed issue of injecting high energy material into the unresolved ISM. It is also most similar to our model, in which we also decouple the wind particles at launch and recouple them once they reach the CGM.

Another recent model is \cite{Costa20}'s which represents the interaction between small scale AGN winds and their larger environment by setting an energy, momentum, and mass flux along a fixed spherical boundary. The boundary itself is built using two rigid, spherical layers of cells. These winds tend to thermalize at distances $\lesssim 50 \, \rm pc$ and produce energy driven bubbles. This model avoids some of the issues faced by existing thermal feedback models such as numerical cooling losses and having feedback decoupled from the instantaneous AGN luminosity (by storing energy for stochastic kicks). The resultant large scale outflows suppress star formation by both removing high density gas from the galaxy nucleus and also reducing the inflow of gas back onto the halo. This provides a good point of comparison for understanding the impact of jets that couple significant amounts of energy or mass into the ISM versus those that do not.

A closely related jet feedback model is \citet{Bourne17}'s, which introduces a hydrodynamic jet prescription in \textsc{AREPO} to study the interaction of AGN jets with the intracluster medium (ICM). 
In this model, mass, momentum, and energy are injected directly into a resolved region around the black hole, allowing the jet to be followed hydrodynamically from launch. A key feature of the \citet{Bourne17} model is the use of specialized refinement and derefinement criteria based on the local jet mass fraction in order to preserve jet structure and limit numerical mixing as the outflow propagates. This is most similar to our refinement methods within \textsc{ArkBH}. Both models emphasize the importance of maintaining high resolution in regions impacted by collimated, high specific energy feedback, but differ in where and how this energy is coupled into the surrounding medium. While their approach couples feedback energy directly into gas cells near the black hole, \textsc{ArkBH} instead uses a hydro-decoupling scheme to transport this energy to larger scales before coupling it to the gas.

A similar approach is taken by \citet{Weinberger17b}, who initialize a collimated, low density outflow within a resolved region a few kpc from the black hole. In their model, resolution within the jet and resulting low density cavities is explicitly increased by imposing a smaller target cell volume, implemented through an adaptive target mass that refines these regions relative to the surrounding medium. In \textsc{ArkBH}, refinement instead serves a different role. Resolution is increased only after feedback energy is deposited in the CGM, using a tracer based scheme that tracks high specific energy outflows and gradually relaxes the resolution in the outflow regions as it propagates. Thus, while both approaches introduce targeted refinement tied to the feedback, they apply it in different ways.

The most significant distinction between \textsc{ArkBH} and other models is the refinement method. \textsc{ArkBH} uses adaptive refinement criteria designed to follow the high specific energy feedback in order to maintain high mass resolution in the region of the CGM traced by the BH jet. With some exceptions, such as \cite{Bourne17} and \cite{Weinberger17b}, many other models rely on the ambient gas resolution. This may be too low to accurately capture the acceleration of hot, collimated outflows, especially within the CGM. Feedback energy in these models may be artificially radiated away (or blow away the ISM entirely), limiting their ability to accurately represent the role of jet driven preventative feedback in quenching galaxies.

\subsection{Caveats}
\label{subsec:Caveats}

We successfully demonstrate that the \textsc{ArkBH} framework performs well in our isolated galaxy, suppressing star formation across most of the explored parameter space. However, there are several limitations to our tests.

First, the above simulations are all run in an isolated Milky Way mass galaxy. Using an isolated galaxy provides a clean environment for development and testing. For that same reason, it lacks many complexities present in cosmological simulations, including galaxy and BH mergers, environmental quenching effects, and larger scale accretion from the IGM. While we do include a gaseous halo to represent the CGM in our isolated setup, our results should thus be interpreted within the context of demonstrating the \textsc{ArkBH} model rather than as a prediction for this halo's evolution. Additionally, the simulations are restricted to running for a maximum of 2~Gyr. While we see some self-regulation due to the feedback, we do not see how the model affects the longer-term galaxy evolution nor the AGN accretion/feedback cycle.

It is also important to note our choice of BH mass. The fiducial BH in our main runs is over massive compared to scaling relations for a Milky Way mass galaxy. The mass was deliberately chosen to ensure strong and sustained feedback, and to test the scheme with a BH mass representative of the masses we expect are most important for AGN feedback. This choice results in an initial burst of high accretion that is not present in our run with an appropriately scaled BH. However, with the lower mass BH run (Fid2), we demonstrate that we can still suppress star formation, and that it is not simply a product of the high initial accretion and associated feedback burst.

Finally, the wind particle launch direction is managed in a simplified way within these simulations. We launch directly along the z-axis, aligned perpendicular to the galactic disk. This ensures that wind particles are launched vertically above or below the disk, matching the predicted biconical behavior of BH jets. This choice is made for simplicity in order to more clearly demonstrate the model. In cosmological simulations, setting the outflow direction is more difficult because galaxies are no longer aligned with the simulation box. There are several alternative ways to set the jet direction, including tracking the BH spin or tying the jet direction to larger scale properties of the galactic disk \citep[e.g.][]{Talbot21, Talbot22, Talbot24, Beckmann24, Huvcko2024}. For example, the launch direction can be set by the cumulative angular momentum of gas accreted onto the BH. Alternative accretion prescriptions (discussed in the next section) can also provide predictions for the accretion disk orientation and therefore inform the outflow direction.

In additional to these missing elements, we emphasize that the current implementation has not used MHD or cosmic rays, both of which may be important in AGN regulation and we plan to include in later versions.

\subsection{Future Directions and Improvements}
\label{subsec:FutureDirections}

We identify several key areas for building upon the existing model as well as future science applications.

One such area is the black hole physics itself. Our simulation currently estimates the BH accretion rate using the Bondi-Hoyle-Lyttleton formalism, capped at the Eddington accretion rate. This is standard across many subgrid models. However, this model makes several simplifying assumptions, including spherical symmetry. More recent accretion prescriptions, such as the torque-driven accretion model \citep[see][]{Hopkins10b, AnglesAlcazar15} or the unified accretion disk model \citep{Koudmani24}, may provide a more physically motivated accretion rate in disk dominated systems. 

Regarding \textsc{ArkBH}'s science applications, we reiterate that a primary focus is to understand the impact of jet driven feedback on gas within the CGM and beyond. The model provides the new resolution techniques required to resolve interactions between high specific energy outflows and gas in these low density regions. We can therefore investigate in depth how the balance of kinetic versus thermal energy as well as the outflow mass loading affects heating and gas inflow/outflow on larger scales. A key direction for future work is the application of \textsc{ArkBH} in cosmological simulations, where these processes can be examined across a statistically representative galaxy population. The future incorporation of \textsc{Arkenstone Cold} will also allow us to investigate multiphase outflows. This combined with our ability to change the outflow opening angle will allow us to extend the model beyond the `jet mode' feedback considered in this work.

A particularly promising observational application of this framework is the study of red geysers, a population of quiescent galaxies characterized by twin nuclear outflows (e.g., Roy et al. 2018, 2021a; Frank 2021). These systems exhibit suppressed star formation despite retaining substantial gas reservoirs, suggesting that AGN driven outflows regulate gas supply without destroying disk structure. By combining \textsc{ArkBH} with the forthcoming \textsc{Arkenstone Cold} framework, we will be able to resolve both the energetic outflow and the embedded cool gas required to potentially reproduce their observed emission-line signatures. Red geysers also therefore provide a direct observational test of whether feedback that primarily couples into the galaxy on larger scales (as in \textsc{ArkBH}), rather than depositing large amounts of energy directly into the ISM, can account for gradual quenching pathways.

This then also brings up the question of how much AGN feedback interacts with the ISM. Early analytical work suggested that AGN feedback coupling to the ISM is important for reproducing observed black hole scaling relations \citep[e.g.][]{Silk98}. This idea is supported by more recent simulations \citep{Hopkins16,Mukherjee18b} that highlight the importance of jet and ISM interactions in matching existing observations. However, not all galaxies exhibit strong evidence for this sort of coupling. Passive spiral galaxies are another example. They show little or no ongoing star formation while retaining their spiral morphology \citep{FraserMcKelvie18}. This implies that quenching can occur without destroying the disk and instead through more gradual or gentle pathways. The \textsc{ArkBH} framework provides a means to explore whether feedback that primarily regulates gas on larger scales rather than depositing large amounts of energy directly into the ISM can better reproduce these systems and their observed properties.

Directly modeling these BH jet/ISM interactions is also an important piece in understanding how BH feedback drives galaxy evolution. Early work by \citet{Sutherland07} and \cite{Wagner12} modeled the impact of jets on an inhomogeneous ISM. They demonstrated different phases of jet evolution: an energy driven bubble phase and a `channel' phase in which the ISM clouds breaks up the jet into streams of plasma entraining ISM material. Later work \citep{Mukherjee16, Mukherjee18} respectively described the impact of relativistic jets on a turbulent two phase ISM and a turbulent gaseous disk. They showed that jets drive turbulence in the ISM, producing a multiphase, filamentary structure. They also showed that high energy jets are less destructive to the ISM, despite driving faster outflows. More recent work has expanded on this, studying jet/ISM interactions at higher resolution to better understand the effects of ISM substructure and turbulence on jet propagation \citep{Ward24, Borodina25}.

While the limited ISM resolution in cosmological simulations makes it very difficult to model the injection of high specific energy feedback into the ISM, the multiphase structure of the \textsc{ArkBH} model allows outflow material to be injected within the ISM when desired. Guided by both our results and existing studies of jet/ISM interactions, future implementations can allocate a fraction of the feedback budget to gas closer to the black hole. These outflows can be parameterized differently than the high specific energy jets. This will allow us to explore how different choices for where and how feedback energy and mass are deposited affect observational signatures and overall galaxy evolution.

\section{Conclusion}
\label{sec:Conclusion}

\textsc{ArkBH} is an extension of the novel \textsc{Arkenstone} method, which was originally designed to resolve high specific energy stellar feedback driven outflows in cosmological simulations. \textsc{ArkBH} adds a framework for launching and resolving black hole outflows, which play an important role in modulating galaxy evolution in massive halos ($M_\text{halo} \gtrsim 10^{12} \, \rm M_\odot$). In this methods paper, we present the additions to the \textsc{Arkenstone} framework designed for BH feedback and again discuss how \textsc{Arkenstone} resolves the injection and evolution of these outflows within the CGM. We use a series of simulations of an isolated $10^{12} \, \rm M_\odot$ galaxy to verify the model, explore the model's parameter space and demonstrate how these outflows influence the host galaxy's evolution. The key results are as follows:

\begin{itemize}
    \item \textbf{Suppressing star formation via preventative feedback:} We demonstrate that ArkBH significantly reduces the cool gas mass and SFR by injecting energy into the CGM. This prevents the inflow of cool gas back onto the ISM and maintains a lower SFR over longer time scales than purely ejective feedback. 

    \item \textbf{Preserved ISM/disk structure:} High resolution simulations indicate jets may punch through the galaxy, leaving the bulk of the ISM largely untouched. \textsc{ArkBH} injects energy directly at the boundary of the ISM and CGM and beyond. By recoupling particles into the CGM, we maintain the ISM/disk structure and avoid significantly disrupting BH accretion. We note that increasing either the total energy injected (i.e. increasing $\epsilon_f$) or specific energy (i.e. increasing $v_\text{w}$) also increases the volume within the recoupling radius that is directly affected by the feedback (i.e. increased temperature, decreased density). This should be taken into consideration when setting the recoupling radius. 

    \item \textbf{Recoupling prescription strongly impacts feedback outcomes:} The density based recoupling (RecRho) produces qualitatively different behavior from fixed distance recoupling by allowing wind particles to recouple within the ISM. This leads to greater suppression of BH accretion due to reduced local gas densities but weaker star formation suppression. This emphasizes that the location of energy deposition has a large impact on the galaxy evolution.
    
    \item \textbf{Suppression of star formation across parameter space:}
    As expected given the large energy injection, \textsc{ArkBH} suppresses star formation across a broad range of parameters, including feedback efficiency, wind velocity, and mass factor. Further, the model remains numerically stable and effective even for extreme configurations, such as velocities an order of magnitude above the typical $10^4 \, \rm km \, s^{-1}$. Increasing the efficiency and wind velocity strengthens suppression but also reduces BH accretion, thereby slowing BH growth.
    
    \item \textbf{Improved refinement strategy:} We successfully demonstrate the hot wind tracer based refinement. The cell mass is lowered in hot wind cells based on the hot wind tracer. This ensures cells with hot wind material have a high enough mass resolution to accurately model the outflow's evolution without having to increase the global mass resolution. This results in outflows that maintain high velocities out to larger radii and propagate farther. Further, we show that outside of our extreme test, $\gamma_\text{m}$ = 1.0, the suppression of star formation in this isolated galaxy is not sensitive to our choice of the mass factor. The tested values produces a similar outflow and galaxy evolution.
\end{itemize}

This work lays the foundation for a full multiphase black hole feedback model within the \textsc{Arkenstone} framework. The next step will be implementing this model in cosmological simulations to further explore the impact of these high specific energy outflows on galaxy evolution. This will also allow us to better understand the impact of the parameter space on the long term evolution of the galaxy and BH. We will then also be able to robustly compare our results against observational work. This includes foundational work showing X-ray cavities and bubbles within the intracluster medium (ICM) \citep{Dunn2006, Dunn2008, Fabian12}. It also includes more recent work measuring ICM velocity dispersions (XRISM), entropy measurements for galaxy groups (SRG/eROSITA), and the kinetic Sunyaev-Zel'dovich effect (DESI). These have already been compared against several existing cosmological simulations \citep{Bahar24, Bigwood25, Xrism25b}.

In future work, we will also implement the second component of \textsc{Arkenstone}, cold cloud modeling, into \textsc{ArkBH}. Adding explicit modeling of unresolved cold cloud evolution is important to capture the full structure and dynamics of AGN driven winds. This will also allow us to model the evolution of the fast molecular outflows observed in galaxies hosting AGN \citep[e.g.][]{Cicone14, Santoro20,Richings18a}. The complete model will again be integrated into cosmological volume simulations, enabling more accurate treatment of galaxy and AGN coevolution.

\section{Acknowledgments}
\label{sec:Acknowledgements}

J.S. acknowledges support through the Frontera Computational Science Fellowship. GLB acknowledges support from the NSF (AST-2108470, AST-2307419), NASA TCAN award 80NSSC21K1053, and the Simons Foundation. J.S.B. acknowledges support from the Leverhulme Trust. S.K. acknowledges support from the Royal Society under grant number URF\textbackslash R1\textbackslash 251867. The simulations and much of the associated data analysis in this work utilized the Frontera and Anvil computing systems at Texas Advanced Computing Center and Purdue University respectively. We thank Rainer Weinberger for comments on the paper and valuable discussion of our results using the TNG BH feedback model. 

We use the Contributor Roles Taxonomy (CRediT)
system\footnote{https://credit.niso.org/} to list the contributions of the co-authors. \textbf{\textit{Conceptualization:}} JS, GB, MS, JB, DF, BT, RS; \textbf{\textit{Data curation:}} JS; \textbf{\textit{Formal analysis:}} JS; \textbf{\textit{Investigation:}} JS;
\textbf{\textit{Methodology:}} JS, GB, MS, JB, DF; \textbf{\textit{Project administration:}} JS, GB; \textbf{\textit{Resources:}} JS, GB, BT; \textbf{\textit{Software:}} JS, MS, JB; \textbf{\textit{Supervision:}} GB; \textbf{\textit{Validation:}}JS, GB, MS, JB, DF, BT; \textbf{\textit{Visualization:}} JS; \textbf{\textit{Writing – original
draft:}} JS; \textbf{\textit{Writing – review $\&$ editing:}} JS, GB, MS, JB, DF, BT, SK, RS, MH. 

\section{Data Availability}
\label{sec:DataAvailability}
The underlying data will be shared upon reasonable request of the corresponding author. 






\bibliographystyle{mnras}
\bibliography{ABH_Bib} 




\FloatBarrier
\appendix

\section{TNG Results}
\label{sec:Appendix}

\begin{figure}
    \centering
    \includegraphics[width=.48\textwidth]{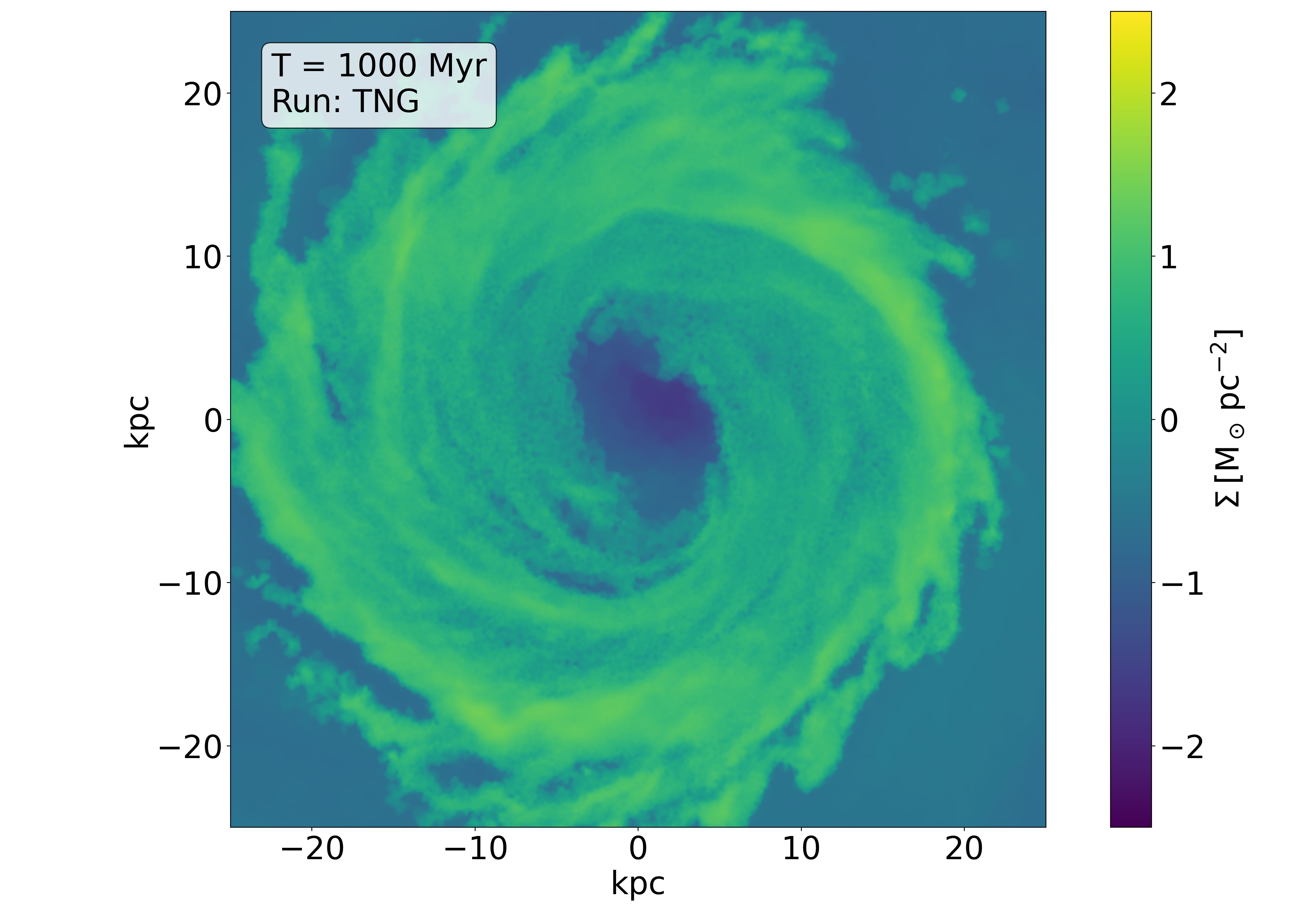}
    \caption{Face-on projection of the gas surface density at t = 1000 Myr in the TNG run.}
    \label{fig:TNGHole}
    
\end{figure}

\begin{figure}
    \centering
    \includegraphics[width=.48\textwidth]{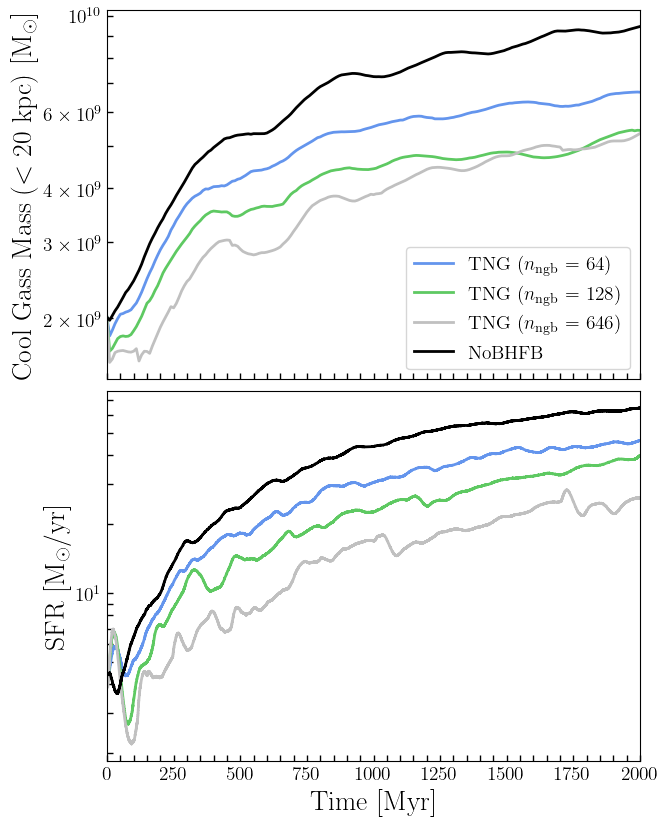}
    \caption{The cool gas mass within 20 kpc of the galaxy center (top panel) and the galaxy's SFR (bottom panel) in several TNG model runs varying $n_\text{ngb}$. We include the original TNG run ($n_\text{ngb}$) and two variations with higher $n_\text{ngb}$ values.}
    \label{fig:TNGSFR}
    
\end{figure}

As mentioned in Section~\ref{subsec:Fiducial}, the TNG BH feedback has a limited effect on the galaxy's SFR. In this run (see Figs.~\ref{fig:FiducialSFR}-\ref{fig:FiducialRays}), the kinetic BH feedback dominates and is responsible for driving material out of the central region of the galaxy. This opens up a large hole in the gas surrounding the BH, reducing the BH accretion rate (Fig.~\ref{fig:TNGHole}). This reduction in the central region's gas density suppresses the galaxy's inner star formation. However, it does not suppress star formation further out in the disk and the reduced black hole accretion slows the future feedback. We enter a quasi-equilibrium state between the BH accretion and feedback, drastically affecting the BH growth. This should not be taken as a general representation of the TNG model, which has been demonstrated to successfully quench galaxies \citep[e.g.][]{Genel18, Weinberger18, Nelson19, Terrazas20}. There are several potential reasons for this discrepancy. \cite{Weinberger17a} shows that the results depend on the choice of $n_\text{ngb}$, which controls both the estimation of local gas properties and the spatial extent of feedback energy injection. We find that increasing $n_\text{ngb}$ from our initial value ($n_\text{ngb} = 64$) does begin to decrease the stellar mass and SFR in our runs but we still do not effectively suppress star formation (Fig.~\ref{fig:TNGSFR}). It is possible that our ICs also play a role in these results, due to the overmassive nature of the BH and resulting high initial accretion rate. Higher resolution also tends to produce larger galaxy stellar masses, SFR densities and several other galaxy population statistics \citep{Pillepich18}. Our mass resolution is most similar to TNG50, which has produced similar results. In particular, \cite{Frosst25} shows this `quasi-equilibrium' style behavior is common among galaxies whose SMBHs experience kinetic feedback. Outside of an initial thermal feedback burst, the BH feedback remains in the kinetic mode, explaining the similarity our results. Finally, we do not include stellar feedback, which is very important for suppressing star formation at this galaxy mass. 

Ultimately, these results show that the TNG feedback in our galaxy regulates gas inflow on the scale of the BH kernel ($\sim \rm kpc)$ rather than the halo or even outer disk scale. Without large scale gas inflow driven by mergers or other effects, we enter the equilibrium state described above.

\section{Mass Factor}
\label{subsubsec:Appendix2}

\begin{figure}
    \centering
    \includegraphics[width=.48\textwidth]{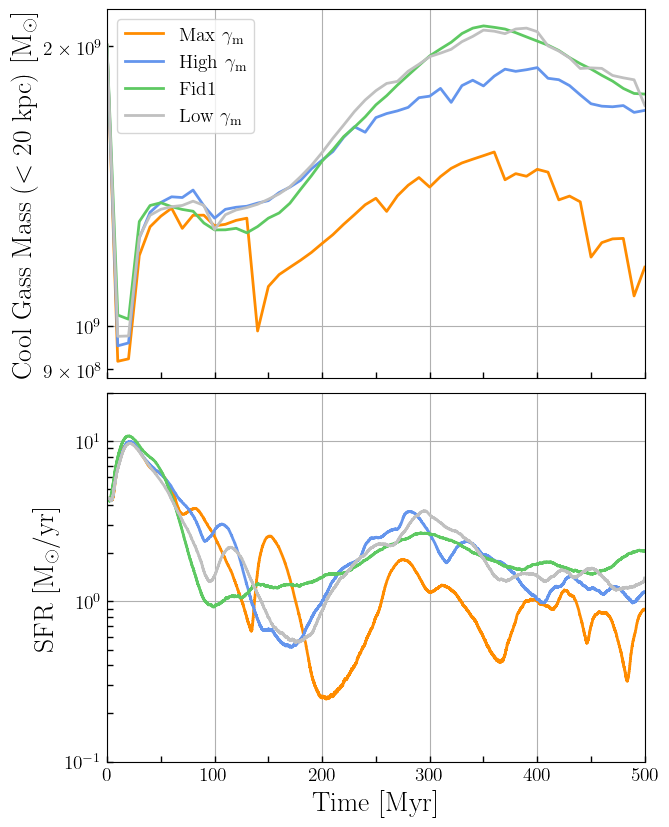}
    \caption{The cool gas mass within 20 kpc of the galaxy center (top panel) and the galaxy's SFR (bottom panel) in the mass factor test runs. The gray, green, blue, and orange lines display the Low $\gamma_\text{m}$, Fid1, High $\gamma_\text{m}$, and Max $\gamma_\text{m}$ runs respectively.}
    \label{fig:MassFactorSFR}
    
\end{figure}

\begin{figure}
    \centering
    \includegraphics[width=.48\textwidth]{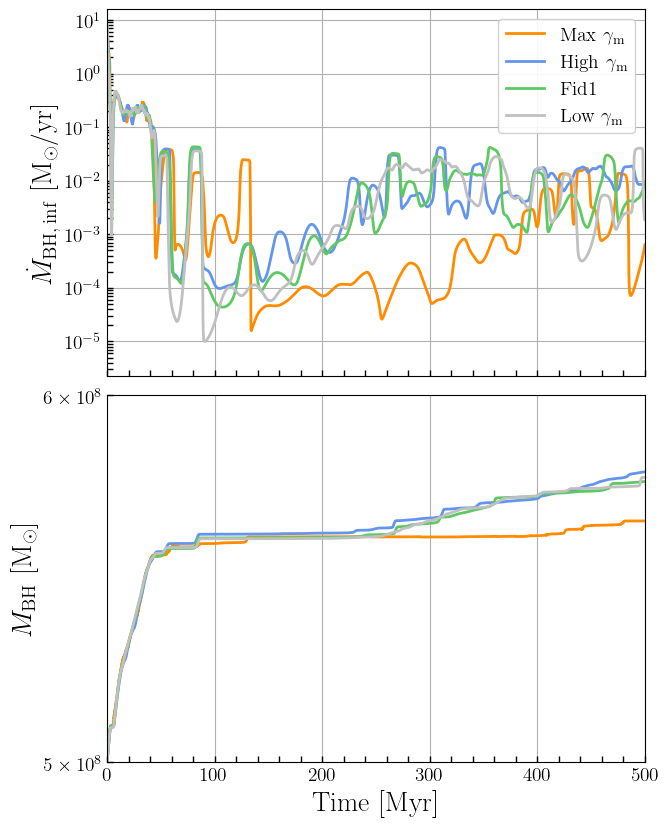}
    \caption{The top and bottom panels display the BH's Bondi accretion rate and total mass, as in Fig.~\ref{fig:FiducialMassAccretion}. The gray, green, blue, and orange lines display the Low $\gamma_\text{m}$, Fid1, High $\gamma_\text{m}$, and Max $\gamma_\text{m}$, as in Fig.~\ref{fig:MassFactorSFR}.}
    \label{fig:MassFactorMassAccretion}
    
\end{figure}

\begin{figure}
    \centering
    \includegraphics[width=.48\textwidth]{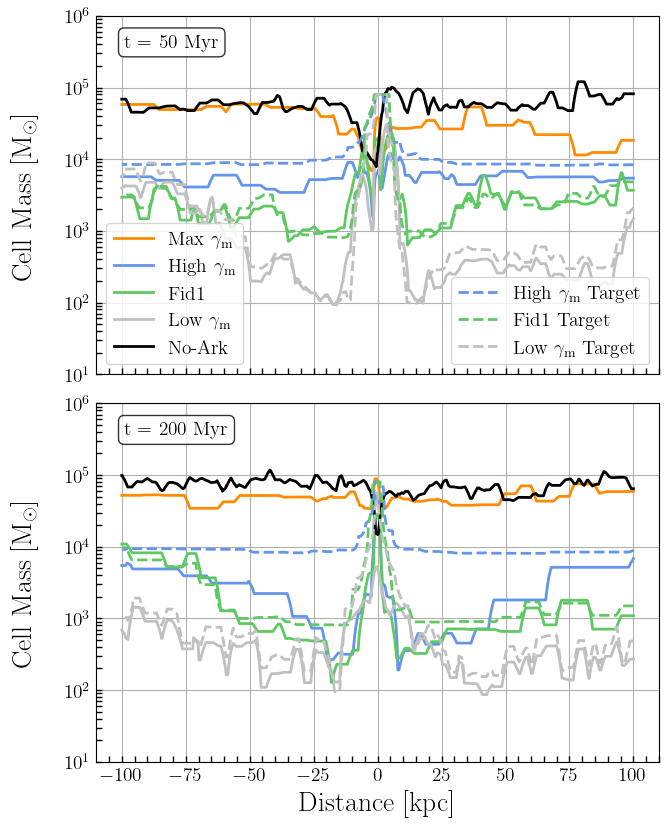}
    \caption{Cell mass profiles for the Low $\gamma_\text{m}$, Fid1, High $\gamma_\text{m}$, Max $\gamma_\text{m}$, and NoBHFB runs at $t = 50\, \rm Myr$ (top) and $t = 200 \, \rm Myr$ (bottom). The dashed cell mass profiles again represent the Arkenstone target cell masses (eq.~\ref{eq:MassRes}). Each profile is measured in the same way as in Fig.~\ref{fig:FiducialRays}. The cell masses differ as expected. We do see the High $\gamma_\text{m}$ cell masses refined above the expected value (cell mass is lower than expected). These cells trigger a flag to avoid de-refining due to the large volume gradient between their neighboring cells.}
    \label{fig:MassFactorRays}
    
\end{figure}

We also investigate the effect of changing the mass factor, $\gamma_{\text{m}}$, which controls the mass resolution of the jet. 

The three mass factor runs between $\gamma_\mathrm{m} = 0.001$–$0.1$ produce very similar effects on the star formation and cool gas mass (Fig.~\ref{fig:MassFactorSFR}). There is a small amount of deviation in the SFRs between the runs, most notably so in the High $\gamma_\text{m}$ run. It has a period of slightly higher star formation at the start of the run. This results in a slightly lower final cool gas mass. However, it does not affect the main results. The difference in mass factor between these runs is not affecting the model's ability to suppress star formation. The Max $\gamma_\text{m}$ run, in which the wind particle masses are set equal to the target gas mass resolution, produces the largest deviation from Fid1. There are fewer but more massive wind particles, producing a `bursty' outflow behavior. This carries over into the cool gas mass and SFR. 

Similarly, we did not observe any significant differences in the BH accretion and final masses across the three $\gamma_\mathrm{m} = 0.001$–$0.1$ runs (Fig.~\ref{fig:MassFactorMassAccretion}). We do see a decrease in the black hole accretion and final mass in the Max $\gamma_\text{m}$ run. This is tied to the sharp decrease in the gas mass after a large feedback burst just before $ t =150 \, \rm Myr$.

Finally, we measured ray profiles of the gas density, temperature, and cell mass in the same way as in the previous runs. The Low $\gamma_\text{m}$, Fid1, and High $\gamma_\text{m}$ runs display a similar density and temperature profile. The largest difference is in the cell masses  (Fig.~\ref{fig:MassFactorRays}). The Low $\gamma_\text{m}$ run generally produces lower cell masses and the High $\gamma_\text{m}$ run produces higher cell masses, both relative to Fid1. We do note that the cell masses in High $\gamma_\text{m}$ are frequently less massive (higher resolution) than what Arkenstone tries to refine them too. The cells do not de-refine to the preferred Arkenstone resolution because they trigger a flag to avoid de-refining if the volume gradient between neighboring cells is too large (greater than a factor of 8 difference). This is good practice to include and not an issue, as we are over-, rather than under-, refined. By comparison, the Max $\gamma_\text{m}$ run has a higher gas density and lower gas temperature within the same region. This implies that some of the injected outflow energy is lost to cooling, as we discuss in Section~\ref{subsec:ArkRecoupling}.

\section{Effects of Arkenstone Refinement}
\label{subsubsec:Appendix3}
We include an additional set of runs to display the effect of the Arkenstone refinement scheme. The first, `ArkBH,' uses both the fiducial set of \textsc{ArkBH} parameters and also the full \textsc{Arkenstone} refinement scheme described in Sec.~\ref{subsec:ArkRefinement}. The second run, `Simple, $\gamma = 0.01$,' uses the same set of parameters but only implements simple recoupling. This means that wind particles recouple into cells on average 100 times more massive than themselves. The final run, `Simple, $\gamma = 1.0$,' again only uses the simple refinement scheme but has an increased $\gamma_{\text{m}}$ value, now set to 1.0. The wind particles in this run are therefore set equal to the target gas mass resolution. 

The first two runs produce have very similar results when focusing on the BH growth and star formation. These similarities are expected and explained in \cite{Smith24a}. Energy losses in the outflow region are negligible due to the high velocities and cooling times. Since the mass and energy loadings are the same across the two runs, there is a similar mass and energy flux. This produces a comparable overall evolution in star formation and the local gas properties that regulate BH growth. 

The third run has a different mass and energy loading due to the change in $\gamma_{\text{m}}$. There are fewer but more massive wind particles, again producing a `bursty' outflow behavior. The BH grows more slowly due to the outflow having a higher mass loading. This run also has the largest impact on the SFR. By both removing the additional refinement and increasing the mass of the particles, we artificially increase the distance around the recoupling point that the injected energy spreads across. This produces a lower SFR due to the outflow energetics `falling back' onto the star-forming disk. The full \textsc{Arkenstone} refinement is therefore needed to accurately capture the propagation of the outflow, which we discuss further. 

\begin{figure}
    \centering
    \includegraphics[width=.48\textwidth]{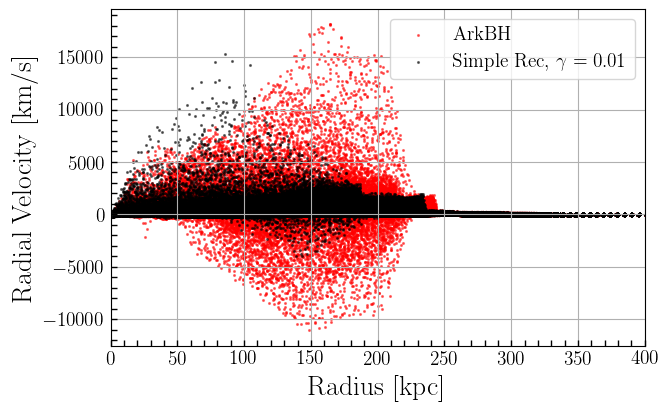}
    \caption{We display a scatter plot of the radial velocities at t = 50 Myr in Fig.~\ref{fig:RefinementZSlices}'s first two runs. `ArkBH' refers to the run with our fiducial set of \textsc{ArkBH} parameters and the Arkenstone refinement scheme, matching the Fid1 run. The `Simple Rec, $\gamma = 0.01$' run uses the same set of parameters but only includes simple recoupling. This means there is no additional refinement applied to the outflow region. }
    \label{fig:RefinementRadialVelocity}
    
\end{figure}

\begin{figure*}
    \centering
    \includegraphics[width=.96\textwidth]{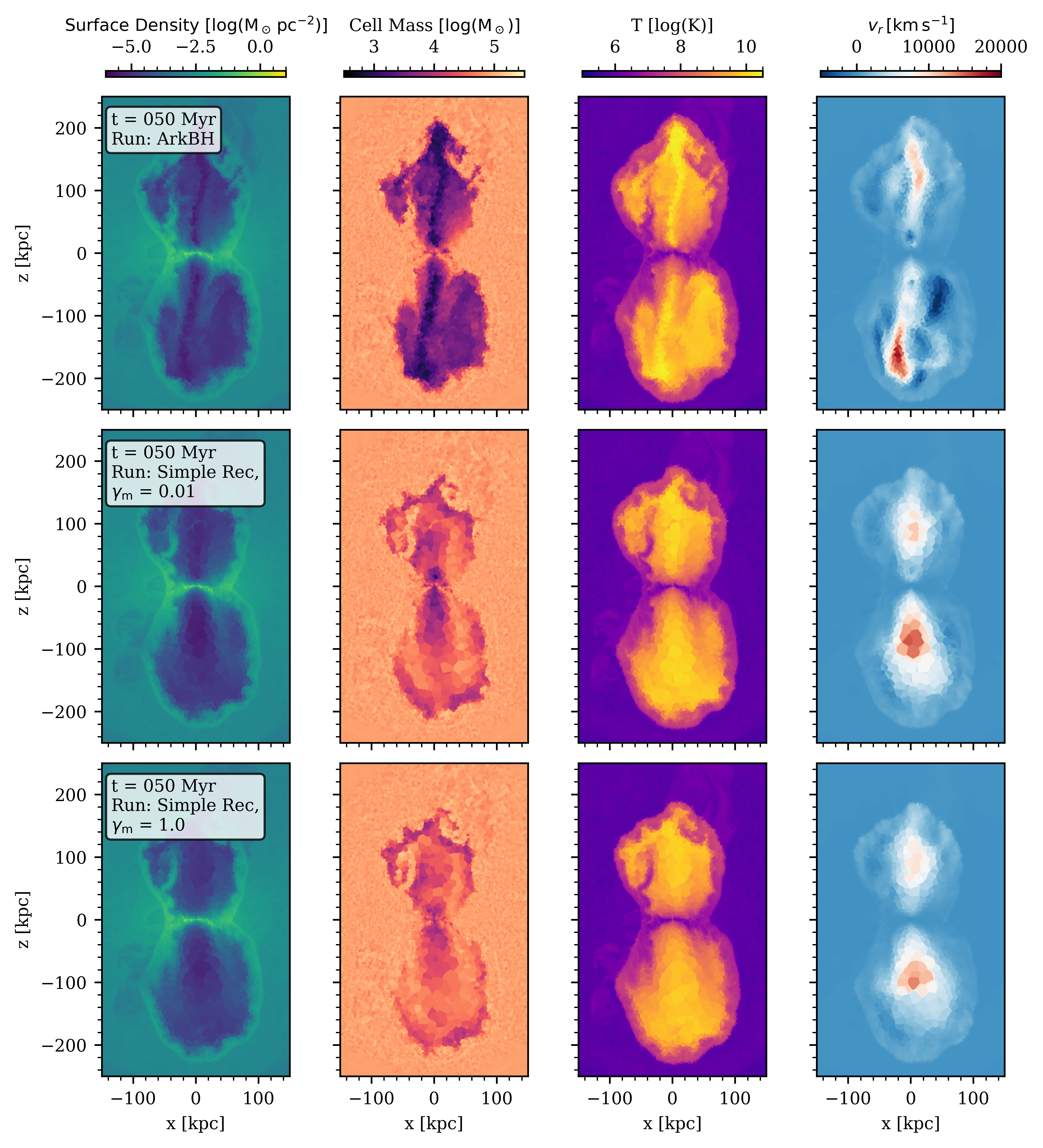}
    \caption{We display three runs from top to bottom: a run with the fiducial set of \textsc{ArkBH} parameters and the Arkenstone refinement scheme described in Sec.~\ref{subsec:ArkRefinement}; a run with the same set of parameters but without the Arkenstone refinement scheme (using only the `simple' refinement scheme from that section); and a second run both with only simple recoupling and also the wind particle masses set to the target gas mass resolution ($\gamma_{\text{m}}$ increased from 0.01 to 1.0). From left to right, we display slices taken along the galaxy's x-z plane of the surface density, cell mass, gas temperature, and radial velocity. These slices are displayed at T = 50 Myr across for each run.}
    \label{fig:RefinementZSlices}
\end{figure*}

With \textsc{Arkenstone}, the outflow maintains higher velocities out to large radii (Fig.~\ref{fig:RefinementRadialVelocity}). In contrast, the outflow radial velocity rapidly falls off at $\sim 100 \, \rm kpc$ in the runs that only include simple recoupling. As \cite{Smith24a} explains, the low resolution without \textsc{Arkenstone} means that the central regions of the outflow where thermal energy should be converted to kinetic energy is unresolved. Further, the injection region is much larger, causing the sonic point to be displaced to larger radii. These two effects combine to produce more retention of thermal energy and less acceleration of gas. The outflow thus extends out further (see Fig.~\ref{fig:RefinementZSlices}) and carries significantly more kinetic energy with it in the ArkBH run. This has significant implications for comparisons to observations. For example, observations of the kinetic Sunyaev–Zel'dovich effect are in tension with many current cosmological simulations, which neither accelerate gas to sufficiently large radii nor inject enough energy into the CGM to match the observations. Properly resolving the acceleration of outflows is therefore an important step toward understanding how feedback produces these signals. The refinement method introduced in \textsc{Arkenstone} shows promise for resolving this outflow of energy and mass out to large radii.

\FloatBarrier


\bsp	
\label{lastpage}
\end{document}